\documentclass[letterpaper]{JHEP3}
\usepackage{amsfonts}

\usepackage{epsfig}

\newcommand{\beq}{\begin{equation}}
\newcommand{\eeq}{\end{equation}}
\newcommand{\beqa}{\begin{eqnarray}}
\newcommand{\eeqa}{\end{eqnarray}}
\newcommand{\beqar}{\begin{eqnarray*}}
\newcommand{\eeqar}{\end{eqnarray*}}

\newcommand{\norm}[1]{\raise.3ex\hbox{:}#1\raise.3ex\hbox{:}}

\newcommand{\zeb}{\zeta_{{\cal B}}}

\abstract{
We consider higher dimensional generalizations of the four
dimensional topological Taub-NUT-AdS solutions,
where the angular spheres $(\theta,\varphi)$ are replaced
by planes and hyperboloids.
The thermodynamics of these configurations is discussed to some extent. The results we find suggest that the
entropy/area relation is always violated
in the presence of a NUT charge. We argue also that the conjectured AdS/CFT
correspondence may teach us something about
the physics in spacetimes containing closed timelike curves.
To this aim, we use the observation that the boundary metric of a
$(D+1)$-dimensional  Taub-NUT-AdS solution
provides a $D$-dimensional generalization of
the known G\"odel-type spacetimes.}
\preprint{hep-th/0407110\\
\newline \small \hfill FREIBURG-THEP-04-06}

\title{Nut Charged Space-times and Closed Timelike Curves on the Boundary}
\author{Dumitru Astefanesei,$^1$\thanks{%
E-mail: \texttt{dastefanesei@perimeterinstitute.ca}} \ Robert B. Mann$^{2}$%
\thanks{%
E-mail: \texttt{mann@avatar.uwaterloo.ca}} \ and Eugen Radu$^3$\thanks{%
E-mail: \texttt{radu@thphys.may.ie}} \\
$^{1}$Department of Physics, McGill University Montr\' eal, Qu\' ebec H3A
2T8, Canada\\
$^{1,2}$Perimeter Institute for Theoretical Physics, Ontario N2J 2W9, Canada%
\\
$^{2}$Department of Physics, University of Waterloo Waterloo, Ontario N2L
3G1, Canada\\
$^{3} $Department of Mathematical Physics, National University of Ireland,
Maynooth, Ireland}
\begin{document}


\section{Introduction}

The anti-de Sitter/Conformal Field Theory (AdS/CFT) correspondence \cite%
{Maldacena:1997re} is the most concrete and astonishing realization of the
holographic principle, which asserts that a consistent theory of quantum
gravity in $(D+1)$ dimensions must have an alternate formulation in terms of
a nongravitational theory in $D$ dimensions. Although string theory in AdS
space is still too complicated to be dealt with in detail, in many
interesting cases, it is sufficient to consider the low energy limit of the
superstring theory, namely, supergravity (see ref.~\cite{Aharony:1999ti} for
a review).

Using this correspondence, one can gain new insights into both theories on
either side of the duality. On the one hand, quantum gravity is reformulated
as an ordinary quantum field theory. This offers the possibility that a full
quantum theory of gravity could be described by a CFT theory that is
understood. Observables in the quantum gravity theory could be computed
using quantum field theory methods. This makes it possible to attempt a
microscopic analysis of the Bekenstein-Hawking entropy of asymptotically AdS
black holes. On the other hand, the proposed duality has become a useful
tool that can be used to learn something about the properties of CFTs that
are only partially understood. Several gauge field theory phenomena such as
confinement, confinement/deconfinement phase transitions and conformal
anomalies have been shown to be encoded in the semiclassical physics of
asymptotically AdS black holes \cite{Aharony:1999ti}.

The duals of AdS quantum gravity solutions are in general CFTs defined on
curved backgrounds with fixed geometries. Thus, supposing the conjecture is
true, we can use it to make some predictions about the basic features of a
CFT in the background of a fixed boundary metric. The geometry of the
boundary of an AdS solution depends on how the spacetime is sliced as one
approaches the boundary, and a large variety of metrics can be obtained
(see, $e.g.$, ref.~\cite{Emparan:1999pm}). We remark that in the cases of
correspondence that we do understand, the asymptotic spacetimes close to the
boundary are globally hyperbolic, with a well defined notion of time. The
field quantization in this case is well understood and, at least in
principle, one can verify the predictions from the gravity side.

It would be interesting to understand how the AdS/CFT correspondence works
in other cases, where the boundary metric presents causal pathologies, in
which case usually there is no well-defined quantum field theory. An
explicit example of such a situation is provided by solutions of the
Einstein equations with negative $\Lambda $ and a nonvanishing NUT charge.
On the supergravity side, there is nothing wrong with considering these
types of configurations since the AdS/CFT correspondence permits
consideration of spacetimes that asymptotically approach AdS space only
locally, but not globally. In the four dimensional case, this corresponds to
the AdS version of the well-known Taub-NUT solution of Einstein's vacuum
equations. This metric represents a generalization of the Schwarzschild
solution \cite{Hawking} and played an important role in conceptual
developments in general relativity. For $\Lambda <0$, it has been shown \cite%
{Chamblin:1999pz} that this spacetime can be generalized by including an
additional discrete parameter $k$ . The case $k\neq 1$ corresponds to
solutions where the angular spheres $(\theta ,\varphi $) are replaced by
planes $(k=0)$ or hyperboloids ($k=-1$). In the limit of vanishing NUT
charge, these solutions correspond to topological black holes (see ref.~\cite%
{Mann:1997iz} for reviews of the subject).

The thermodynamics of the four dimensional solutions has been discussed by
several authors with some surprising features (see, $e.g.$, refs.~\cite%
{Emparan:1999pm, Chamblin:1999pz}). A discussion of $(D+1)$-dimensional 
generalizations of the $k=1$ configurations~\cite%
{Clarkson:2002uj} employed the Euclidean approach, continuing
the time and NUT charge to imaginary values. In this case the mass parameter
has to be restricted such that the fixed point set(s) of the Euclidean time
symmetry is (are) regular. In four dimensions one obtains in this way two
types of solutions, ``nut'' or ``bolt'', depending on whether the fixed
point set has zero dimensionality (a ``nut'') or is of $2$ dimensions (a
``bolt''). In $(D+1)$ dimensions one can not only have fixed point sets of
co-dimension $(D-1)$ (bolts) or zero (nuts) but also cases of intermediate
dimensionality \cite{Clarkson:2002uj,Mann:2003zh}.

The thermodynamics of the $k=1$ solutions depends on the number of spacetime
dimensions. One of their more interesting features is that the entropy is
not proportional to the area. The origin of this phenomenon has generally
been regarded as being due to the fixed point sets of lower
co-dimensionality (the nuts) which manifest themselves as Misner strings.
For nut solutions in $(4p+2)$ dimensions (with $p=1,2,...$) there are no
regions in parameter space for which the entropy and specific heat are both
positive definite, suggesting a thermodynamic instability. However, all bolt
solutions have some regions of parameter space for which thermodynamic
stability is possible, $i.e.$ both entropy and specific heat are positive
definite. To our knowledge, a similar discussion of the higher dimensional
generalizations of the $k=0,-1$ Taub-NUT-AdS solutions is still missing in
the literature.

We argue in this paper that NUT charged spacetimes are important from yet
another viewpoint: namely we may hope to learn something about
quantum field theories in nonglobally hyperbolic backgrounds. As noted in
ref.~\cite{us} (see, also, ref.~\cite{Radu:2002hf}), the boundary metrics of
a four dimensional Lorentzian Taub-NUT-AdS spacetime correspond to the
essential three-dimensional part of a homogeneous G\"{o}del-type spacetime,
and contain closed timelike curves (CTCs) for a large class of parameters %
\cite{reboucas}. We recall that a G\"{o}del-type spacetime is the archetypal
cosmology exhibiting properties associated with the rotation of the
universe. Additionally, the G\"{o}del model \cite{Godel:1949ga} is perhaps
the best known example of a solution of Einstein's field equations in which
causality may be violated. It thus became a paradigm for causality violation
in gravitational theory. Recently there has been renewed interest in this
kind of spacetime because of the discovery of a supersymmetric solution of $%
N=1$ five dimensional supergravity with very similar features \cite%
{Gauntlett:2002nw}. This five dimensional solution possesses many
interesting properties: it has proven to be dual to certain $pp-$waves \cite%
{Boyda:2002ba,Brecher:2003rv} and it can be extended so as to embed a black
hole within it \cite{Gimon:2003ms}. In this context, Boyda et al. \cite%
{Boyda:2002ba} claimed that the chronology is well defined once a G\"{o}%
del-type spacetime is prescribed with macroscopic holography: their proposal
is to keep a part of the causal region associated with a comoving observer,
replacing the rest with an observer dependent screen (see, also, ref.~\cite%
{nov} for an early attempt in this direction).

Motivated by these results, we consider in this paper higher dimensional
generalizations of the $k=0,-1$ topologically Taub-NUT-AdS solutions and
discuss their thermodynamic properties. We find the surprising result that
these solutions have an entropy that is not proportional to the area,
despite the absence of Misner strings \cite{Astefanesei:2004ji}. We
also find that the first law of thermodynamics forces the temperature to be
inversely proportional to the NUT charge; only in the $k=1$ case
is the constant of proportionality forced to take on integer values so the
Misner string singularities are removed. Furthermore, we find that although
these solutions have non-zero angular momentum, this makes no contribution
to the partition function because the angular velocity of the horizon
vanishes. By studying the boundary metric of Taub-NUT-AdS spacetimes, we are
led to a natural generalization in higher dimensions of the known
homogeneous G\"{o}del-type spacetimes. Not surprisingly, the line element of
the five-dimensional G\"{o}del-type supersymmetric solution found in ref.~%
\cite{Gauntlett:2002nw} is the boundary of a six-dimensional $k=0$
topological Taub-NUT-AdS solution.

The paper is structured as follows: in the next section we obtain the $(D+1)-
$dimensional general solution of the Einstein equations with negative
cosmological constant and nonzero NUT charge, which generalizes the known
four dimensional configurations. In Section 3 we discuss its thermodynamic
properties on the Euclidean section. We show that the breakdown of the
entropy/area relationship does not depend upon the removal of singularities
associated with Misner strings. As discussed in Section 4, the boundary
metric of the general solution presented in Section 1 gives a higher
dimensional generalization of the known G\"{o}del-type line elements with
many similar properties. In Section 5 we compile the results, proposing a
preliminary discussion of the implications of the AdS/CFT correspondence for
the physics in a causality violating background. Appendix A shows
that Misner string singularities are absent in the $k=0,-1$ cases
and appendix B contains a brief discussion of the zeta function 
regularization method and the Euclidean effective action for a scalar 
field propagating in a G\"{o}del-type line element, explicit results 
being presented in a particular case.

Most of the notation and sign conventions used in this paper are similar to
those in ref.~\cite{Clarkson:2002uj}.


\section{A general solution}

In this context, we start by considering $(D+1)$-dimensional vacuum
Taub-NUT-AdS solutions, with a negative cosmological constant $\Lambda
=-D(D-1)/2\ell^{2}$. Here $D$ is of the form $D=2m+1$, with $m$ a positive
integer. The general solution in four dimensions (which can be obtained by
analytic continuation of the Euclidean solutions discussed in ref.~\cite%
{Chamblin:1999pz}) is 
\begin{eqnarray}  \label{metric-4D}
ds^{2}=\frac{dr^{2}}{F(r)}+(r^{2}+n^{2})\Big(d\theta ^{2}+f_{k}^{2}(\theta
)d\varphi ^{2}\Big)-F(r)\Big(dt+4nf_{k}^{2}(\frac{\theta }{2})d\varphi \Big)%
^{2},
\end{eqnarray}
where 
\begin{eqnarray}  \label{F}
F=k\left( \frac{r^{2}-n^{2}}{r^{2}+n^{2}}\right) +\frac{-2Mr+\frac{1}{\ell
^{2}}(r^{4}+6n^{2}r^{2}-3n^{4})}{r^{2}+n^{2}}.
\end{eqnarray}%
The discrete parameter $k$ takes the values $1,0,-1$ and implies the form of
the function $f_{k}(\theta )$ 
\begin{eqnarray}  \label{f}
f_{k}(\theta )=\left\{ 
\begin{array}{ll}
\sin \theta , & \mathrm{for}\ \ k=1 \\ 
\theta , & \mathrm{for}\ \ k=0 \\ 
\sinh \theta , & \mathrm{for}\ \ k=-1.%
\end{array}%
\right.
\end{eqnarray}%
The properties of the case $k=1$ are well known from the $\Lambda =0$
Taub-NUT solution \cite{NUT,Misner}. In the limit $\Lambda \rightarrow 0$,
this metric has become renowned for being \emph{``a counterexample to almost
anything''} \cite{misner-book} and represents a nontrivial generalization of
the Schwarzschild solution \cite{Hawking}. It is usually interpreted as
describing a gravitational dyon with both ordinary and magnetic mass. The
NUT charge $n$ plays a role dual to that of the ordinary mass $M$, in the
same way that electric and magnetic charges are dual within Maxwell theory %
\cite{dam}.

As discussed by many authors, the presence of magnetic-type mass (the NUT
parameter $n$) introduces for $k=1$ a ``Dirac-string singularity'' in the
metric (but no curvature singularity). This can be removed by appropriate
identifications and changes in the topology of the spacetime manifold, which
imply a periodic time coordinate (and thus CTCs). A similar analysis yields
the same pathologies for a $k=1$ spacetime with a negative cosmological
constant. However, generalizing the previous analysis for $k=1$ \cite{Misner}, 
one can prove that, for $k\neq 1$, there are
no Misner strings even if the $\left( \theta ,\varphi \right)$ section is 
compact (see Appendix A). Thus, there is no requisite periodicity
of the time coordinate $t$, although in general there are not
reasonable foliations by spacelike surfaces for any value of $k$.

In fact, it can directly be proven that, for a nonzero NUT charge, a
negative cosmological constant implies the possible existence of CTCs in the
bulk for a certain range of the parameters.  To prove that, we consider the
curve generated by the Killing vector $\partial /\partial \varphi $ and
study the quantity 
\[
g_{\varphi \varphi }=4f_{k}^{2}(\frac{\theta }{2})\left(
r^{2}+n^{2}-f_{k}^{2}(\frac{\theta }{2})(4n^{2}F+k(r^{2}+n^{2}))\right) . 
\]%
One can see that, for all $k=1,0$ metrics and those $k=-1$ metrics with $%
4n^{2}/\ell ^{2}>1$, the curve $r=r_{0},~t=$const$.,~\theta =\theta _{0}\neq
0$ becomes timelike for large enough values of $(r_{0},~\theta _{0})$,
corresponding to a CTC since $\varphi $ is a periodic coordinate. This type
of causality violation is familiar from the study of the G\"{o}del
spacetime. There are also regions of these spacetimes which are free of CTCs.

The $k=-1$ metrics with $4n^2/\ell^2\leq 1$ are globally hyperbolic, $%
f(x^i)=t$ being a global time coordinate ($i.e.$ $g^{ij}(\partial f/\partial
x^i) (\partial f/\partial x^j) <0$ everywhere). In the limit $n \to 0$, the $%
k=0,-1$ metrics correspond to topological black holes and have been
considered in AdS/CFT context by many authors (see, $e.g.$, ref.~\cite%
{emparan}).

Higher dimensional generalizations of the line element (\ref{metric-4D}) in
the $k=1,~\Lambda =0$ case were first considered by Bais and Batenberg \cite%
{Bais:1984xb}. A negative cosmological constant was introduced by Awad and
Chamblin \cite{Awad:2000gg}, also only for $f_{1}(\theta )=\sin \theta $.
While there are a number of papers dealing with the properties of the $k=1$
solutions with NUT charge \cite{Hawking:1998ct}, the cases $k=0,-1$ have
received relatively little attention (however, see ref.~\cite{Mann:2003zh}).

The general form for the Lorentzian Taub-NUT-AdS class of metrics for a $%
U(1) $ fibration over $(M^{2})^{\otimes p}$ is 
\begin{eqnarray}  \label{TN-ADS-gen}
ds^{2}=\frac{dr^{2}}{F(r)} +G(r)(d\theta _{i}^{2}+f_k^{2}(\theta_i) d\varphi
_{i}^{2}) -F(r)(dt +4n f_k^2( \frac{\theta _{i}}{2}) d\varphi _{i})^2,
\end{eqnarray}
with $i$ summed from $1$ to $p$. The base space $M^{2}$ corresponds to a two
dimensional sphere $(k=1)$, plane $(k=0)$ or pseudohyperboloid $(k=-1$).

By solving the Einstein equation in $(D+1)$ dimensions with a negative
cosmological constant, we find $G(r)=r^{2}+n^{2}$, while the general form
for $F(r)$ is found to be 
\begin{eqnarray}
F(r)=\frac{r}{(r^{2}+n^{2})^{(D-1)/2}}\Big (\int^{r}\left[ k\frac{%
(s^{2}+n^{2})^{(D-1)/2}}{s^{2}}+\frac{D}{\ell ^{2}}\frac{%
(s^{2}+n^{2})^{(D+1)/2}}{s^{2}}\right] ds-2M\Big),
\end{eqnarray}%
where the parameter $M$ is an integration constant related to the spacetime
mass. These spacetimes contain only quasiregular singularities, which are
the end points of incomplete and inextensible geodesics that spiral
infinitely around a topologically closed spatial dimension. Moreover the
world lines of observers approaching these points come to an end in a finite
proper time \cite{Konkowski}. The Riemann tensor and its derivatives remain
finite in all parallelly propagated orthonormal frames, and no curvature
scalars diverge, making them the mildest kind of singularity.

On the Lorentzian section, the parameters $M$ and $n$ are no longer related
but they can be freely specified (for $k=-1$, $M$ may take negative values).
The roots of $g_{tt}=-F(r)$ correspond to horizons, while the horizon region
is always nonsingular. Again, requiring the absence of singularities due to
the Misner-string implies for $k=1$ a periodicity $2\pi (D+1)n$ of the time
coordinate. For $k=0,-1$ the fibration is trivial: there are no Misner
string singularities and no periodicity of $t$. However, similar to the four
dimensional case, $t$ is a global time coordinate for those $k=-1$ metrics
with $2(D-1)n^{2}/l^{2}<1$ only. Also, all $k=1,0$ metrics and those $k=-1$
configurations with $4n^{2}/\ell ^{2}>1$ have $g_{\varphi _{i}\varphi _{i}}<0
$ for some range of $r,\theta _{i}$ (thus there are CTCs extending to
infinity). In the limit $n=0$ we find the black hole solutions ~\cite%
{Mann:1997iz} 
\[
ds^{2}=\frac{dr^{2}}{F(r)}+r^{2}\left( d\theta _{i}^{2}+f_{k}^{2}(\theta
_{i})d\varphi _{i}^{2}\right) -F(r)dt^{2},~~~\mathrm{with}~~~F(r)=\frac{k}{%
D-2}-\frac{2M}{r^{D-2}}+\frac{r^{2}}{\ell ^{2}},
\]%
with $M=0$ corresponding to a special foliation of the AdS$_{D+1}$ spacetime.


\section{Euclidean TNAdS and thermodynamic properties}

For solutions with NUT charge, most of the calculations in the literature
are carried out for the Euclidean section. The usual approach to quantum
gravity in this case is to analytically continue in the time coordinate and
the NUT charge \cite{Hawking:ig} 
\begin{eqnarray}  \label{cont}
t\rightarrow i\tau ,~~n\rightarrow iN,
\end{eqnarray}%
in order to obtain a solution on the Euclidean section of the Einstein
equations (with negative cosmological constant). Thus, the Euclidean section
of the general Taub-NUT-AdS solution (\ref{TN-ADS-gen}) reads 
\begin{eqnarray}  \label{E-TN-ADS-gen}
ds^{2}=\frac{dr^{2}}{F(r)}+(r^{2}-N^{2})(d\theta _{i}^{2}+f_{k}^{2}(\theta
_{i})d\varphi _{i}^{2})+F(r)(d\tau +4Nf_{k}^{2}(\frac{\theta _{i}}{2}%
)d\varphi _{i})^{2},
\end{eqnarray}%
where now 
\begin{eqnarray}
F(r)=\frac{r}{(r^{2}-N^{2})^{(D-1)/2}}\Big (\int^{r}\left[ k\frac{%
(s^{2}-N^{2})^{(D-1)/2}}{s^{2}}+\frac{D}{\ell ^{2}}\frac{%
(s^{2}-N^{2})^{(D+1)/2}}{s^{2}}\right] ds-2M\Big),
\end{eqnarray}%
and the radial coordinate is greater than the largest root of $F(r)$. The
function $F(r)$ can be written in a more transparent way 
\begin{eqnarray}
F(r) &=&\frac{r^{2}-N^{2}}{\ell ^{2}}-\frac{2Mr}{(r^{2}-N^{2})^{(D-1)/2}}+(k-%
\frac{N^{2}(D+1)}{\ell ^{2}})  \label{Fe2} \\
&&\times \frac{r}{(r^{2}-N^{2})^{(D-1)/2}}\Big (\int^{r}ds\frac{%
(s^{2}-N^{2})^{(D-1)/2}}{s^{2}}\Big ).  \nonumber
\end{eqnarray}%
The properties of the resulting configurations in four dimensions have been
extensively discussed by a number of authors \cite%
{Emparan:1999pm,Hawking:1998ct}. For example it was found that these
Euclidean solutions cannot be exactly matched to AdS at infinity.

There is an important feature that distinguishes the solutions with $k=1$
from solutions with $k=0,-1$. The $k=1$ Euclidean solutions have the
coordinate $\tau $ identified with the period 
\begin{equation}
\beta =\left| \frac{2\pi N}{\sigma }\right| (D+1),  \label{temp}
\end{equation}%
which is determined by demanding regularity of the manifold so that the
singularities at $\theta _{i}=0,\pi $ are coordinate artifacts ($\sigma$ is 
a constant, see below). This
property is not shared by $k=0,-1$ solutions since there are no
Misner strings in these cases (see Appendix A). However we will find
that the self-consistency of the thermodynamic relations forces eq. (\ref%
{temp}) to hold for these two cases as well.

For any value of $k$, the absence of conical singularities at the roots $%
r_{+}$ of the function $F(r)$ imposes a periodicity in the Euclidean time
coordinate 
\begin{equation}
\beta =\left| \frac{4\pi }{F^{\prime }(r_{+})}\right| .  \label{new-rel}
\end{equation}%
Usually, $\beta $ is interpreted as the inverse temperature. For $k=1$, the
parameter $\sigma $ must be integer-valued so that the periodicity in $\tau $
induced by eq. (\ref{temp}) is commensurate with the periodicity induced by
eq. (\ref{new-rel}). No such restriction on $\sigma $ holds for the $k=0,-1$
cases.

For any $k$, the mass parameter $M$ must be restricted such that the fixed
point set of the Killing vector $\partial _{\tau }$ is regular at the
radial position $r=r_{+}$ (defined by setting $F(r_{+})=0$). We find in this
way two types of solutions, ``bolt'' (with arbitrary $r_{+}=r_{b}>N$) or
``nut'' $(r_{+}=N)$, depending on whether the fixed point set is of
dimension $(D-1)$ or is less than this maximal value.

The Lorentzian solutions (\ref{TN-ADS-gen}) extremize the gravitational
action \cite{Gibbons:1976ue} 
\begin{eqnarray}  \label{action}
I=I_{B}+I_{\partial B},
\end{eqnarray}%
where 
\begin{eqnarray}  \label{actbound}
I_{B} &=&-\frac{1}{16\pi G}\int_{\mathcal{M}}d^{D+1}x\sqrt{-g}\left(
R-2\Lambda \right) , \\
I_{\partial B} &=&-\frac{1}{8\pi G}\int_{\mathcal{\partial M}}d^{D}x\sqrt{%
-\gamma }\, \Theta .
\end{eqnarray}%
The first term in eq.~(\ref{action}) is the bulk action over the $(D+1)$%
-dimensional manifold $\mathcal{M}$ with metric $g$ and the second term (\ref%
{actbound}) is the surface term necessary to ensure that the Euler-Lagrange
variation is well-defined. Here, $\gamma $ is the induced metric of the
boundary and $\Theta$ is the extrinsic curvature.

The gravitational action computed in this way (even at tree-level) contains
divergences that arise from integrating over the infinite volume of
spacetime. In the AdS/CFT context, the infrared (IR) divergences of gravity
are interpreted as ultraviolet (UV) divergences of the dual CFT. A well
understood way\footnote{%
This method works, also, for some asymptotically non anti-de Sitter spaces
(see, $e.g.$, ref.~\cite{noAdS} and references therein).} of computing the
bulk action without introducing a background is to add local counterterms
into the action \cite{Balasubramanian:1999re}, which remove all divergences,
leading to a finite action corresponding to the partition function of the
CFT. Thus we have to supplement eq.~(\ref{action}) with 
\begin{eqnarray}
I_{\mathrm{ct}} &=&\frac{1}{8\pi G}\int d^{D}x\sqrt{-\gamma }\left\{ -\frac{%
D-1}{\ell }-\frac{\ell \mathsf{\Theta }\left( D-3\right) }{2(D-2)}\mathsf{R}-%
\frac{\ell ^{3}\mathsf{\Theta }\left( D-5\right) }{2(D-2)^{2}(D-4)}\left( 
\mathsf{R}_{ab}\mathsf{R}^{ab}-\frac{D}{4(D-1)}\mathsf{R}^{2}\right) \right.
\nonumber  \label{Lagrangianct} \\
&&+\frac{\ell ^{5}\mathsf{\Theta }\left( D-7\right) }{(D-2)^{3}(D-4)(D-6)}%
\left( \frac{3D+2}{4(D-1)}\mathsf{RR}^{ab}\mathsf{R}_{ab}-\frac{D(D+2)}{%
16(D-1)^{2}}\mathsf{R}^{3}\right.  \nonumber \\
&&\left. -2\mathsf{R}^{ab}\mathsf{R}^{cd}\mathsf{R}_{acbd}\left. -\frac{D}{%
4(D-1)}\nabla _{a}\mathsf{R}\nabla ^{a}\mathsf{R}+\nabla ^{c}\mathsf{R}%
^{ab}\nabla _{c}\mathsf{R}_{ab}\right) +...\right\} ,
\end{eqnarray}%
where $\mathsf{R}$ and $\mathsf{R}^{ab}$ are the curvature and the Ricci
tensor associated with the induced metric $\gamma $. The series truncates
for any fixed dimension, with new terms entering at every new odd value of $%
D $, as denoted by the step-function ($\mathsf{\Theta }\left( x\right) =1$
provided $x\geq 0$, and vanishes otherwise).

Using these counterterms, one can construct a divergence-free boundary
stress tensor from the total action $I=I_{B}+I_{\partial B}+I_{ct}$ by
defining a boundary stress-tensor 
\[
T_{ab}=\frac{2}{\sqrt{-h}}\frac{\delta I}{\delta h^{ab}}, 
\]%
whose explicit expression for $D\leq 8$ is given in ref.~\cite{Das:2000cu}.
Thus a conserved charge 
\begin{equation}
\mathfrak{Q}_{\xi }=\oint_{\Sigma }d^{D-1}S^{a}~\xi ^{b}T_{ab},
\label{Mcons}
\end{equation}%
can be associated with a closed surface $\Sigma $ (with normal $n^{a}$),
provided the boundary geometry has an isometry generated by a Killing vector 
$\xi ^{\mu }$ \cite{Booth}. If $\xi =\partial /\partial t$ then $\mathfrak{Q}
$ is the conserved mass/energy $\mathfrak{M}$. The angular momentum $J_{i}$
along the $i$ direction is usually associated with the Killing vector $\Psi
=\partial /\partial \varphi _{i}$ --- although for the metric form (\ref%
{TN-ADS-gen}), a more natural choice is $\partial /\partial \varphi
_{i}-2kN\partial /\partial t$ \cite{Misner}.

We can proceed further by formulating gravitational thermodynamics via the
Euclidean path integral 
\[
Z=\int D\left[ g\right] D\left[ \Psi \right] e^{-I\left[ g,\Psi \right]
}\simeq e^{-I}, 
\]%
where one integrates over all metrics and matter fields between some given
initial and final Euclidean hypersurfaces, taking $\tau $ to have some
period $\beta $, generally determined by demanding regularity of the
Euclideanized manifold at degenerate points of the foliation.
Semiclassically the result is given by the classical action evaluated on the
equations of motion, and yields to this order an expression for the entropy 
\begin{equation}
S=\beta (\mathfrak{M}-\mu _{i}\mathfrak{C}_{i})-I,  \label{GibbsDuhem}
\end{equation}%
upon application of the Gibbs-Duhem relation to the partition function \cite%
{Mann:2003 Found} (with chemical potentials $\mathfrak{C}_{i}$ and conserved
charges $\mu _{i}$). The specific heat $C$ is computed with the standard thermodynamic formula 
\begin{equation}
\label{C-gen}
C=-\beta \frac{\partial S}{\partial \beta }. 
\end{equation}%
In ref.~\cite{Clarkson:2002uj}, the action, mass/energy, entropy and
specific heat of the spherically $(D+1)-$dimensional Taub-NUT-AdS solutions
were calculated in this way.


\subsection{The four dimensional case}

We start by applying this procedure to the better-known four dimensional
case.

The $k=1$ solutions have an intrinsic Euclidean time periodicity $\beta
=8\pi N$, fixed by demanding the absence of Dirac string singularities. For $%
k=0,-1$ the fibration is trivial and the periodicity of $\beta $ is not
fixed by this requirement. Detailed analyses of the $k=1$ nut and bolt
solutions were presented in refs.~\cite{Emparan:1999pm, Chamblin:1999pz,
Clarkson:2002uj} and we shall not repeat them here.

The mass/energy of the topological TNAdS solutions calculated by the
counterterm formalism is $\mathfrak{M}=VM/(4\pi G)$, which yields the
familiar $D=3$ expression for the topological black hole mass in the limit $%
N\rightarrow 0$ (in this section, $V$ is the dimensionless area of
the $\theta ,\varphi $ space, and can be normalized to $4\pi $).

Interestingly enough, since asymptotically 
\[
T_{\varphi t}=\frac{8Mnf_{k}^{2}(\theta /2)}{\ell r}+O(1/r^{2}),~~T_{tt}=%
\frac{2M}{\ell r}+O(1/r^{2}), 
\]%
these solutions also have a nonzero angular momentum density, while the
total angular momentum is obtained by computing eq.~(\ref{Mcons}). 
This quantity vanishes for the $k=1$ solution, but yields a nonzero value $%
J\sim nM$ in the other two cases.

To compute the associated angular velocity $\Omega $ entering the
thermodynamics, we consider a family of locally nonrotating observers ``at
rest'' with respect to the $t=const.$ hypersurface \cite{Wald:rg}, with
velocity $u^{a}\sim \nabla ^{a}t$ . These observers rotate with an angular
velocity $\Omega =-g_{t\varphi }/g_{\varphi \varphi }$. Unlike the Kerr-AdS
case, this vanishes as $r\rightarrow r_{+}$ . Thus the contribution of $J$
to the thermodynamic partition function is zero.\footnote{%
A similar situation has been found for some asymptotically flat rotating
black holes in an Einstein-Maxwell-dilaton theory \cite{Kleihaus:2003df}.}

Nut solutions have fixed point sets at $r=N$. The mass parameter of planar
nut solutions ($k=0$) must take the value $M_{n}=-4N^{3}/\ell ^{2}$, while
the metric function $F(r)$ takes the form 
\[
F(r)=\frac{(r-N)^{2}(r+3N)}{\ell ^{2}(r+N)},
\]%
with a double zero at $r=N$. As a result, these solutions correspond to a
background with zero temperature \cite{Chamblin:1999pz}, since the
Euclidean time $\tau $ can be identified with any period $\beta $. The
action and the entropy of this solution are 
\begin{eqnarray}
I_{0,4}^{N}=-\frac{V\beta }{4\pi G}\frac{N^{3}}{\ell ^{2}}, ~~~~
S_{0,4}^{N}=- \frac{3V\beta}{4\pi G}\frac{ N^{3}}{\ell^{2}}.  \nonumber
\end{eqnarray}
It can be verified that the first law of thermodynamics $dS=\beta d 
\mathfrak{M}$ is satisfied, since we can always take $\beta=\sigma N$, with $%
\sigma$ an arbitrary constant.

The situation with $k=0$ bolt solutions is different. Here, $\partial _{\tau
}$ has a two-dimensional fixed point set at some radius $r=r_{b}>N$. In this
case the mass parameter is 
\[
M_{0,4}^{b}=\frac{1}{2\ell ^{2}r_{b}}(r_{b}^{4}-6N^{2}r_{b}^{2}-3N^{4}), 
\]%
the period $\beta $ resulting from eq.~(\ref{new-rel}) is
\[
\beta =\frac{4\pi \ell ^{2}}{3}\frac{r_{b}}{r_{b}^{2}-N^{2}}. 
\]%
while the action and entropy  are
\begin{eqnarray}
I_{0,4}^{b} &=&-\frac{V(r_{b}^{4}+3N^{4})}{12G(r_{b}^{2}-N^{2})},
\\
\label{asa1}\nonumber
S_{0,4}^{b} &=&\frac{V(r_{b}^{4}-4N^{2}r_{b}^{2}-N^{4})}{4G(r_{b}^{2}-N^{2})}.
\end{eqnarray}%
Note that, for any value of $r_{b}$, the entropy is not proportional to the
area $A=V\left( r_{b}^{2}-N^{2}\right) $ of the bolt. This is quite distinct
from the result obtained in computing the entropy of the planar $k=0$ bolt
metric relative to its nut counterpart -- upon appropriate matching of the
geometries at large radius, one obtains a value of $A/4$ for the relative
entropy \cite{Chamblin:1999pz}. However, we see that the counterterm approach
yields an entropy that is not proportional to the area despite the absence
of Misner strings.

The situation in the hyperbolic case is somewhat special. In this case there
are no nuts \cite{Chamblin:1999pz} since $F(r)$ becomes negative for values
of $r$ close to (but larger than) $N$. We find only bolts at some value of $%
r=r_{b}>N$. The corresponding mass parameter is 
\[
M_{-1,4}^{b}=\frac{1}{2\ell ^{2}r_{b}}\left( r_{b}^{4}-(6N^{2}+\ell
^{2})r_{b}^{2}-N^{2}(3N^{2}+\ell ^{2})\right) , 
\]%
while the inverse temperature is given by 
\[
\beta =\frac{4\pi \ell ^{2}r_{b}}{3(r_{b}^{2}-N^{2})-\ell ^{2}}, 
\]%
which is a positive quantity, as follows from the assumptions we made on
the function $F(r)$. For this value of $\beta$, the expressions
of the action and entropy are
\begin{eqnarray}
I_{-1,4}^{b} &=&-\frac{V\left( r_{b}^{4}+\ell ^{2}r_{b}^{2}+N^{2}(\ell
^{2}+3N^{2})\right) }{4G(3r_{b}^{2}-\ell ^{2}-3N^{2})},   
\\
\nonumber
S_{-1,4}^{b} &=&\frac{V\left( \ell ^{2}N^{2}+3N^{4}+\ell
^{2}r_{b}^{2}+12N^{2}r_{b}^{2}-3r_{b}^{4}\right) }{4G(\ell
^{2}+3N^{2}-3r_{b}^{2})}. \label{asa2}
\end{eqnarray}%
\newpage
\setlength{\unitlength}{1cm}

\setlength{\unitlength}{1cm}

\begin{picture}(6,6)
\centering
\put(-1.75,0){\epsfig{file=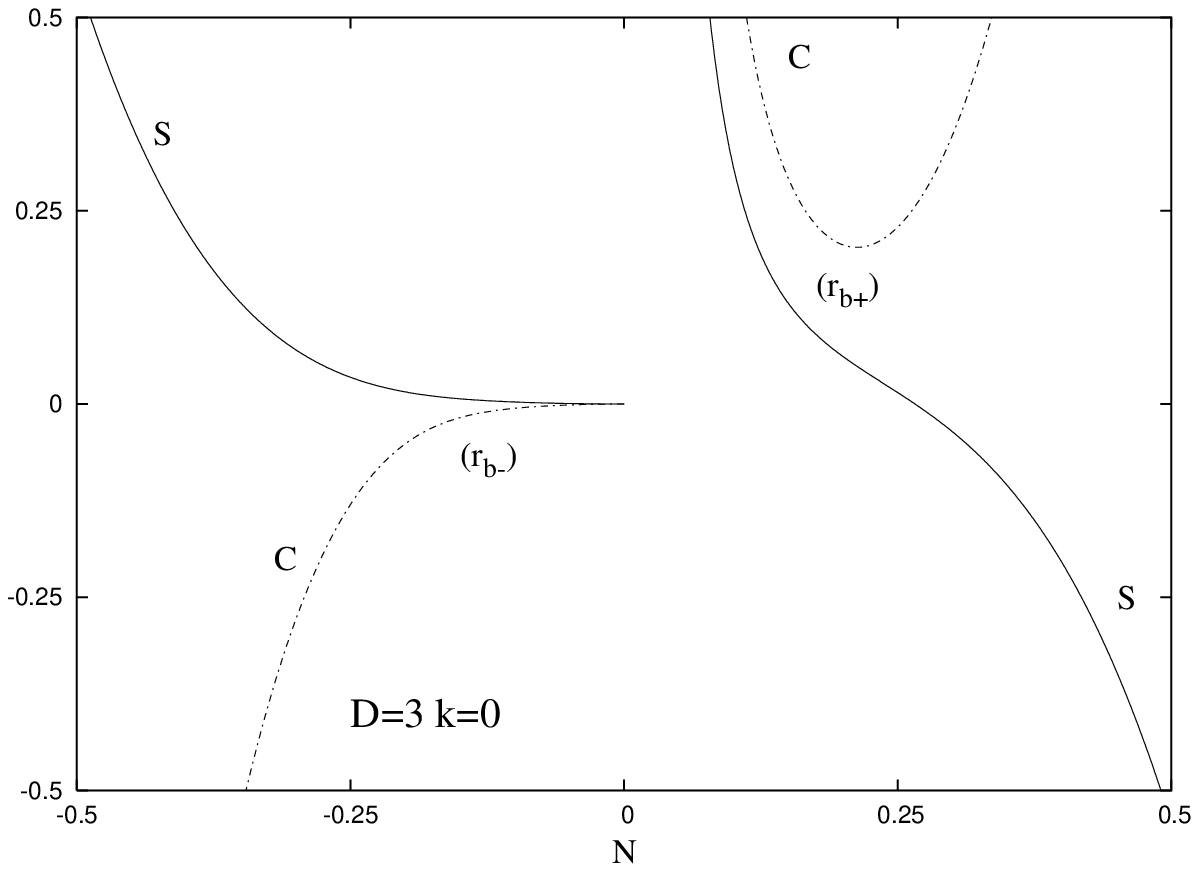,width=8.5cm}}
\end{picture}%
\begin{picture}(-10,1.5)
\centering
\put(0.75,0){\epsfig{file=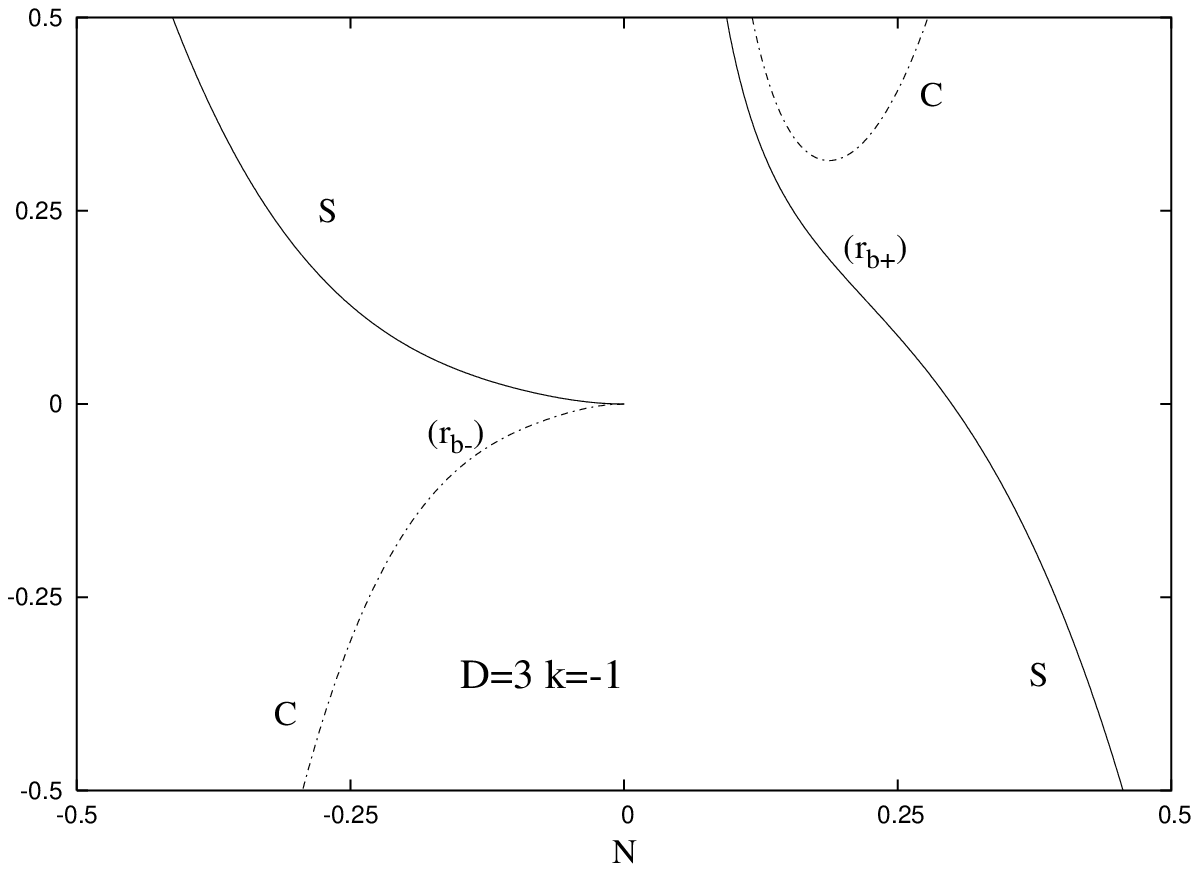,width=8.5cm}}
\end{picture}\newline
{\small \textbf{Figure 1.} The entropy and specific heat of $k=0,-1$
solutions are represented as a function of $N$ for bolt solutions in four
dimensions. }
\\
\\
We can see that, given the arbitrariness in the choice of $r_{b}$, these
quantities may take any values. However, for any $k$, the bolt solutions
with large enough values of $r_{b}$ have $S_{k,4}>0$ and are thermally
stable $C_{k,4}^{b}>0$ (although the action is negative).

For generic $r_{b}>N$, we find that the first law of thermodynamics
is not satisfied for any value of $k$. This suggests that the bolt
radius is not independent of the NUT
charge and/or the cosmological constant. Supposing $r_{b}=r_{b}(N)$ we
find that the first law of thermodynamics holds if and only if
\begin{equation}
r_{b\pm }=\frac{\sigma\ell ^{2}\pm \sqrt{\sigma^{2}\ell
^{4}+48N^{2}(3N^{2}-k\ell ^{2}})}{12N} ,  \label{r-bolts}
\end{equation}
where $\sigma$ is an arbitrary parameter. Insertion of this relationship
into eq. (\ref{new-rel}) implies $\beta =2\pi \left( D+1\right) N/\sigma $.
Remarkably, we find that a consistent thermodynamics forces a relation
between the NUT charge and the bolt radius, despite the absence of Misner
string singularities in the $k=0,-1$ cases. For the $k=1$ spacetimes, these
can be removed by requiring $\sigma $ to be an integer. Insertion of eq. (%
\ref{r-bolts}) into the expression for the entropy yields a value that is
not proportional to the area of the bolt unless $N=0$.

There are two branches that yield consistent thermodynamics. The obvious
condition $r_b^2>N^2$ implies that the lower branch $r_{b-}$ exists for a
negative $N$ only and arbitrary $\sigma$. For $N>0$, the $r_{b+}$ branch
exists when $\sigma>2k$, whereas for a negative $N$, it exists for $\sigma<2k$.

The thermodynamic quantities of these solutions are found by replacing eq.~(\ref%
{r-bolts}) in the general relations (\ref{asa1}), (\ref{asa2}).
For example, the expression of the specific heat  is  
\begin{eqnarray}
C=\frac{V\ell^4}{144 G N^2 \sigma^2} \pm
\frac{V\big(\ell^8-24 \ell^4 N^2 \sigma^2(k\ell^2 -6 N^2)+10368 N^6\sigma^4(-k\ell^2+4N^2) \big) }
{144G \ell^2 N^2 \sigma^2
\sqrt{\ell^4+48 N^2\sigma^2(-k\ell^2 +3 N^2)}}.
\end{eqnarray}
 However, the
arbitrariness of $\sigma $ makes their analysis somewhat difficult \ and the
analytical expressions are not very illuminating.

Figure 1 shows a plot of the entropy and specific heat per unit volume vs. $N$ (for $%
\ell =1$). We see that for small $N$, the $k=0,-1$ upper branch solutions ($%
r_{b}=r_{b+}$) are thermally stable, whereas the lower branch solutions ($%
r_{b}=r_{b-}$) are always unstable. The upper branch quantities diverge in
the high temperature limit ($i.e.$, $N\rightarrow 0$) while 
the lower branch quantities are all zero.
These results are obtained for $|\sigma |=1.5$. Similar results have been
obtained for other values of $\sigma $.


\subsection{The general case}

Now we turn to the general case. For $k=1$, an expression for the mass,
temperature and finite gravitational action in arbitrary dimensions was
derived in ref.~\cite{Clarkson:2002uj}.

Not unexpectedly, the general methods presented there to find the finite
values of the action and mass are generally valid for any value of the
discrete parameter $k$. Therefore, we repeat here only the basic arguments
put forward in Section 7 of ref.~\cite{Clarkson:2002uj}.

In the general case, the bulk contribution to the total action is
\[
I_{B}=\frac{DV\beta }{8\pi G\ell ^{2}}\int_{r_{+}}^{r^{\prime
}}\, dr(r^{2}-N^{2})^{(D-1)/2},
\]%
where $V=\int \prod_{i=1}^{p}f_{k}(\theta _{i})d\theta _{i}d\varphi $ is the
total area of the sector $(\theta _{i},\varphi _{i})$ (with $V=(4\pi
)^{(D-1)/2}$ for $k=1$).\footnote{%
In the non-compact case, $V$ will diverge. If we compactify, then it will be
finite. However, in the divergent case, we can consider entropy and action
per unit $V$.} The integrand here can be expanded using the binomial
theorem. Integrating term by term and taking the limit $r^{\prime
}\rightarrow \infty $, we find that the general expression for the finite
contribution from the bulk action is
\begin{equation}
\label{Ibulkfinite}
I_{B\mathrm{finite}}=-\frac{DV \beta }{8\pi G \ell ^{2}}
\sum_{i=0}^{(D-1)/2}\left( {{\frac{(D-1)}{2} }\atop{i }} \right)
(-1)^{i}N^{2i}%
\left[ \frac{r_{+}^{D-2i}}{D-2i}\right],
\end{equation}
with $r_{+}$ the value of $r$ that is the largest positive root of $F\left(
r\right) $. The expression for the trace of the extrinsic curvature $\Theta $
corresponding to the metric (\ref{E-TN-ADS-gen}) is
\[
\Theta =\frac{F^{\prime }(r)}{2\sqrt{F(r)}}+(D-1)\frac{r\sqrt{F(r)}}{%
(r^{2}-N^{2})}. 
\]%
Thus, expanding $\sqrt{-\gamma }\Theta $ for large $r$, the general finite
contribution from the boundary action is 
\[
I_{\partial B\mathrm{finite}}=\frac{DV\beta }{8\pi }M. 
\]%
To evaluate $I_{ct}$ we make use of the expression of the Ricci scalar of
the $D-$dimensional boundary metric 
\[
\mathsf{R}=(D-1)\Big(\frac{k}{(r^{2}-N^{2})}-\frac{F(r)N^{2}}{%
(r^{2}-N^{2})^{2}}\Big). 
\]%
Using that relation and the expressions for the Ricci tensor of the boundary
metric, we find that for every $k$, the only finite contribution from the
counterterm action (\ref{Lagrangianct}) comes from the first term as $%
r\rightarrow \infty $. All of the other terms will either go to zero or
diverge, and will be used to cancel the divergences in $I_{B}$ and $%
I_{\partial B}$.

Thus, the general expression for the finite gravitational action in $(D+1)$
dimensions is similar to that found in ref.~\cite{Clarkson:2002uj}
\begin{eqnarray}  \label{action-gen}
I_{k,D+1}=\frac{V \beta}{8 \pi G}\left[M
-\frac{D}{l^2}\sum_{i=0}^{(D-1)/2}%
\left( {{\frac{(D-1)}{2} }\atop {i }} \right)
(-1)^{i}N^{2i}\frac{r_{+}^{(D-2i)}}{%
(D-2i)}\right].
\end{eqnarray}
We can also derive a general expression for the mass for the TNAdS
solutions, by using the general relation (\ref{Mcons}). The divergence
cancellations take place in a similar manner and we find for the conserved
mass-energy 
\begin{eqnarray}  \label{MassGen}
\mathfrak{M}=\frac{(D-1)VM}{8\pi G}.
\end{eqnarray}%
We have checked the validity of the results (\ref{action-gen}) and (\ref%
{MassGen}) for $D=3,5,7$.

However, the value of $M$ is fixed by the condition $F(r=r_{+})=0$ to be 
\begin{eqnarray}
M_{k,D+1} &=&\frac{1}{2}\int^{r_{+}}\frac{ds}{s^{2}}\left( (k(s^{2}-N^{2})^{%
\frac{D-1}{2}}+\frac{D}{\ell ^{2}}(s^{2}-N^{2})^{\frac{(D+1)}{2}}\right)
\label{mass-gen} \\
&=&\frac{1}{2}\left( \frac{(r_{+}^{2}-N^{2})^{(D+1)/2}}{l^{2}r_{+}}+(k-\frac{%
N^{2}(D+1)}{l^{2}})\int^{r_{+}}ds\frac{(s^{2}-N^{2})^{(D+1)/2}}{s^{2}}%
\right) .  \nonumber
\end{eqnarray}%
To derive the periodicity $\beta $ of the Euclidean time coordinate, we use
the relations 
\begin{equation}
\left| \frac{dF(r)}{dr}\right| _{r=N}=\frac{2k}{N(D+1)},  \label{F-prim-nut}
\end{equation}%
while for bolt solutions 
\begin{equation}
\left| \frac{dF(r)}{dr}\right| _{r=r_{b}}=\frac{1}{r_{b}}\left( k+\frac{D}{%
\ell ^{2}}(r_{b}^{2}-N^{2})\right) .  \label{Fprim-bolt}
\end{equation}%
The thermodynamics of the general solution can also be discussed to some
extent. For $k\neq 1$, the behavior known from four dimensions is found to
be generic: there are no hyperbolic nuts with $\Lambda <0$, the $k=0$ nuts
have zero temperature and the first law of thermodynamics implies that $%
r_{b} $ is a function of $\ell$ and $N$.

Furthermore, the $(D+1)$ dimensional solutions also have $p$ nonvanishing
charges $J_{i}$, associated with the Killing vectors $\varphi _{i}$ that are
interpreted as angular momenta. However, as with the $D=3$ case, the
angular velocities $\Omega _{i}$ entering the thermodynamic relations are
zero and so the entropy is still given by $S=\beta \mathfrak{M}-I$. 

\subsubsection{General nut solutions}

We start by discussing the simpler case when the largest zero of the
function $F(r)$ is at $r_{+}=N$. Since the properties of the $k=1$ solutions
are discussed in ref.~\cite{Clarkson:2002uj}, we focus here on the cases $%
k=0,-1$. Here we remark that, as implied by eq.~(\ref{F-prim-nut}), $%
dF/dr|_{r=N}=0$ for $k=0$ and any $D$, implying that $F(r)$ has a double
zero as $r_{+}=N$. Thus, the $k=0$ NUT solutions correspond to a background
of arbitrary temperature since the Euclidean time $\tau $ can be identified
with arbitrary period. \ By using the same relation, we can prove that the
``no hyperbolic nuts'' result found in ref.~\cite{Chamblin:1999pz} for $D=3$
also generalizes to higher dimensions with negative cosmological constant.
Eq. (\ref{F-prim-nut}) implies that the function $F(r)$ necessarily has a
second zero for some $r_{+}>N$, and $F(r)$ is positive for $r>r_{+}$.

Making use of the relation 
\[
\int dx\frac{(x^{2}-1)^{n}}{x^{2}}\Big|_{x=1}=-\frac{1}{\sqrt{\pi }}\Gamma (%
\frac{1}{2}-n)\Gamma (n+1), 
\]%
we can write simple expressions for the parameter $M$, action and entropy of
$k=0,1$ NUT solutions 
\begin{eqnarray}
M_{k,D+1}^{N} &=&\frac{\sqrt{\pi }}{2\ell ^{2}}(-1)^{(D-1)/2}N^{D-2}\left(
N^{2}(D+1)-kl^{2}\right) \frac{\Gamma (\frac{D+1}{2})}{\Gamma (\frac{D}{2})},
\nonumber  \label{gen1-3} \\
I_{k,D+1}^{N} &=&\frac{V\beta (-1)^{(D-1)/2}}{16\sqrt{\pi }G\ell ^{2}}N^{D-2}%
\Big((D-1)N^{2}-k\ell ^{2}\Big)\frac{\Gamma (\frac{D+1}{2})}{\Gamma (\frac{D%
}{2})}, \\
S_{k,D+1}^{N} &=&\frac{V\beta (-1)^{(D-1)/2}}{16\sqrt{\pi }Gl^{2}}N^{D-2}%
\Big(D(D-1)N^{2}-(D-2)k\ell ^{2}\Big)\frac{\Gamma (\frac{D+1}{2})}{\Gamma (%
\frac{D}{2})},  \nonumber
\end{eqnarray}%
where $\beta =2\pi N(D+1)$ for $k=1$ and is arbitrary for $k=0$. For $k=1$,
we recover the relations presented in ref.~\cite{Clarkson:2002uj}. In this
case, for any $D$, there is no region in parameter space for which the
entropy and specific heat are both positive definite. Again, for $k=0$, the
first law of thermodynamics is satisfied by taking the periodicity of the
Euclidean time coordinate to be $\beta =\sigma N$, with $\sigma $ an
arbitrary constant. 

\subsubsection{General bolt solutions}

Bolt solutions exist for every value of $k$. In this case, the condition $%
F(r_{b}>N)=0$ implies that 
\[
\beta =\frac{4\pi r_{b}\ell ^{2}}{k\ell ^{2}+D(r_{b}^{2}-N^{2})} 
\]%
is the periodicity of $\tau $ coordinate. The parameter $M_{k,D+1}^{b}$ is
fixed by eq.~(\ref{mass-gen}), with $r_{+}=r_{b}$. Note the existence for $%
k=1$ of an intrinsic periodicity $\beta =\left| \frac{2\pi (D+1)N}{p}\right| 
$, which fixes the value of $r_{b}$ as a function of $(\ell ,N)$.

The action and entropy of the bolt solutions are given by the general
expressions 
\begin{eqnarray}
I_{k,D+1}^{b} &=&\frac{V\beta }{16\pi G\ell ^{2}}\left( -\frac{%
(r_{b}^{2}-N^{2})^{(D+1)/2}}{r_{b}}+\big(k\ell ^{2}-(D-1)N^{2}\big)%
\int^{r_{b}}ds\frac{(s^{2}-N^{2})^{(D-1)/2}}{s^{2}}\right) ,
\label{rel-bolt} \\
S_{k,D+1}^{b} &=&\frac{V\beta }{16\pi G\ell ^{2}}\left( \frac{%
D(r_{b}^{2}-N^{2})^{(D+1)/2}}{r_{b}}+\big((D-2)k\ell ^{2}-D(D-1)N^{2}\big)%
\int^{r_{b}}ds\frac{(s^{2}-N^{2})^{(D-1)/2}}{s^{2}}\right) .  \nonumber
\end{eqnarray}%
We omit here the expression of the specific heat since it is extremely
complicated and not particularly enlightening.

As found in ref.~\cite{Clarkson:2002uj}, all $k=1$ bolt solutions have some
regions of parameter space for which thermodynamic stability is possible. As
expected, this property is shared by all $k=0,-1$ counterparts. The
existence in this case of three essential parameters ($r_b,N,\ell)$ makes
this fact always possible.

We present the expression of the relevant quantities in six dimensions 
\begin{eqnarray}
M_{0,6}^{b} &=&\frac{1}{2\ell ^{2}r_{b}}\left(
r_{b}^{6}-5N^{2}r_{b}^{4}+15N^{4}r_{b}^{2}+5N^{6}\right) ,  \nonumber
\label{k=0;-bolt-6D} 
\\
I_{0,6}^{b} &=&\frac{V\left(
-3r_{b}^{6}+5N^{2}r_{b}^{4}+15N^{4}r_{b}^{2}+15N^{6}\right) }{%
60G(r_{b}^{2}-N^{2})},  
\\
\nonumber
S_{0,6}^{b} &=&\frac{V\left(
3r_{b}^{6}-13N^{2}r_{b}^{4}+33N^{4}r_{b}^{2}+9N^{6}\right) }{%
12G(r_{b}^{2}-N^{2})}, 
\end{eqnarray}%
for planar bolts, and
\begin{eqnarray}
M_{-1,6}^{b} &=&\frac{1}{6\ell ^{2}r_{b}}\left( 3r_{b}^{6}-(15N^{2}+\ell
^{2})r_{b}^{4}+(45N^{4}+6\ell ^{2}N^{2})r_{b}^{2}+15N^{6}+3\ell
^{2}N^{4}\right) ,  
\nonumber  \label{k=-1;-bolt-6D} 
\\
I_{-1,6}^{b} &=&\frac{V\left(
-3r_{b}^{6}+(5N^{2}-l^{2})r_{b}^{4}+3N^{2}(5N^{2}+2\ell
^{2})r_{b}^{2}+3N^{4}(5N^{2}+\ell ^{2})\right) }{12G(r_{b}^{2}-N^{2})},
\\
S_{-1,6}^{b} &=&\frac{V\left( 15r_{b}^{6}-(65N^{2}+3\ell
^{2})r_{b}^{4}+3N^{2}(55N^{2}+6\ell ^{2})r_{b}^{2}+9N^{4}(5N^{2}+\ell
^{2})\right) }{12G(r_{b}^{2}-N^{2})},
\nonumber
\end{eqnarray}%
for the six-dimensional hyperbolic bolts. One can see that, for large enough
values of $r_{b}$ the action is negative, while the mass parameter and entropy
 become positive.

However, as with the four dimensional case, the first law of thermodynamics
is satisfied provided $r_{b}$ is a function of $N$, which in turn implies
that eq.~(\ref{temp}) holds.

 We find the existence of two branches, given by
\[
r_{b\pm }=\frac{\sigma\ell ^{2}\pm \sqrt{\sigma^{2}\ell ^{4}
+D(D+1)^{2}N^{2}\left( DN^{2}-k\ell ^{2}\right) }}{D(D+1)N}.
\]
Unlike from the case $k=1$, the parameters $N,\ell$ may take arbitrary
values. Several restrictions appear from the condition $r_b>N$. For the
upper branch $r_{b+}$, a positive $N$ implies values of $\sigma>k(D+1)/2$,
while, for a negative $N$, we find the condition $\sigma<k(D+1)/2$. 

The lower branch exists for an arbitrary $\sigma$ but negative values of $N$
only.

We have looked for the thermodynamic quantities in a number of dimensions up
to $D=9$ and found a general picture that resembles the $D=3$ case. For
small $N$, the $k=0,-1$ upper branch solutions are thermally stable, whereas
the lower branch solutions are always unstable. The upper branch quantities 
diverge in the high temperature limit
while the lower branch quantities vanish in the same limit. 

\newpage
\setlength{\unitlength}{1cm}

\begin{picture}(6,6)
\centering
\put(-1.75,0){\epsfig{file=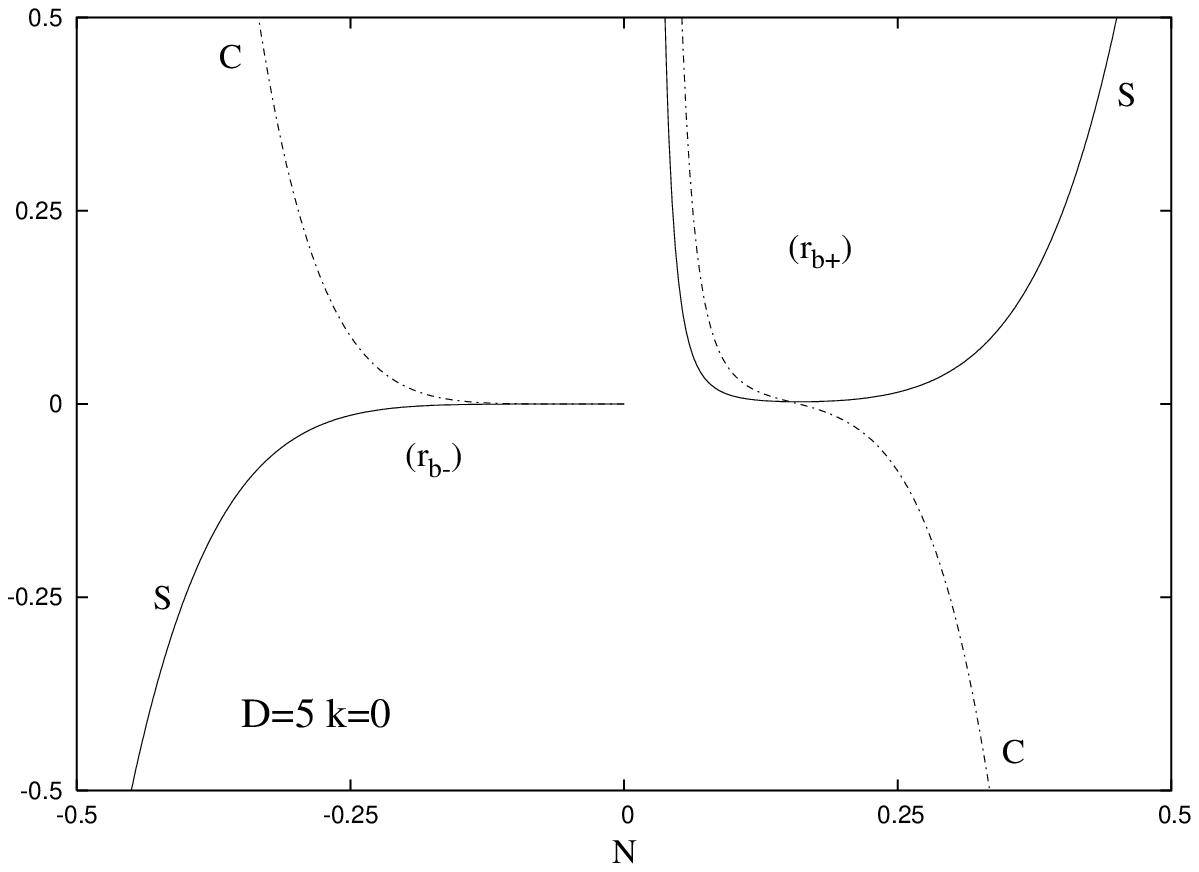,width=8.5cm}}
\end{picture}
\begin{picture}(-10,1.5)
\centering
\put(0.75,0){\epsfig{file=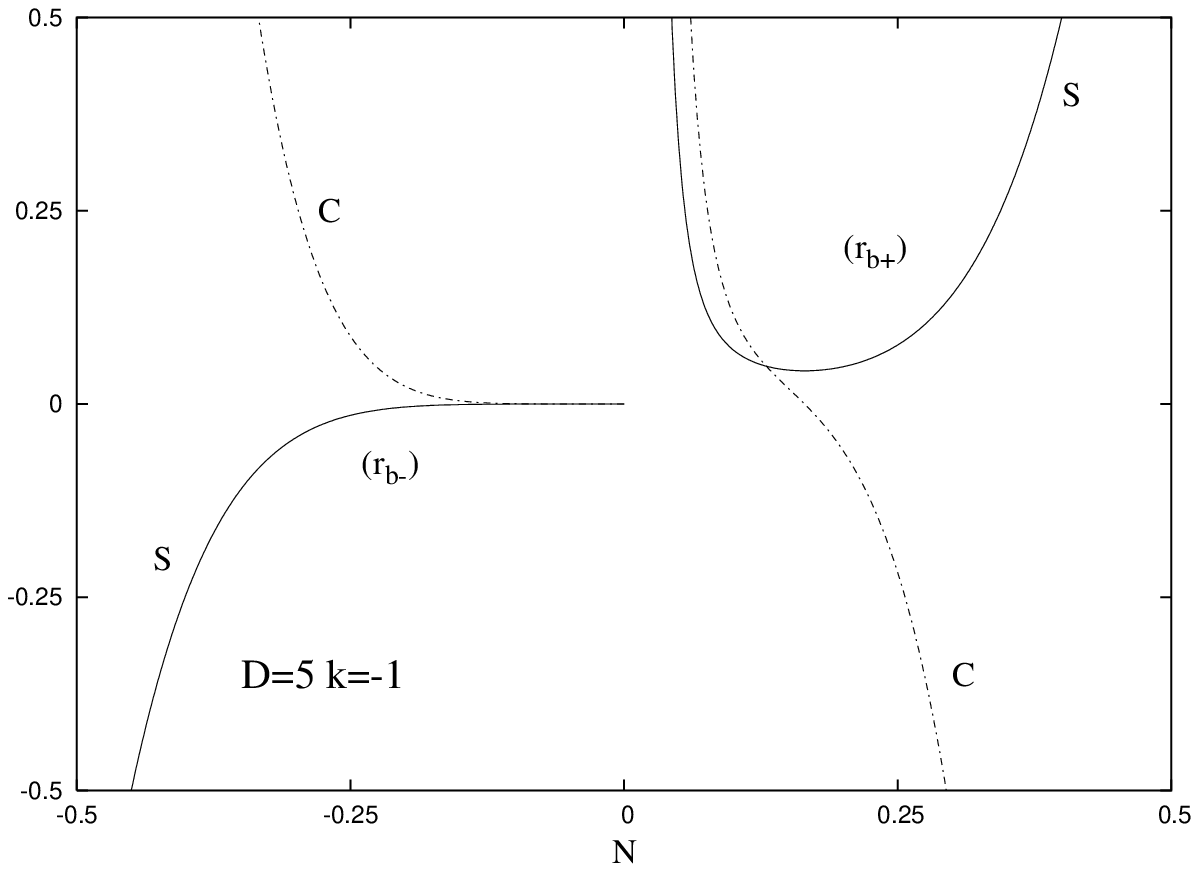,width=8.5cm}}
\end{picture}
\newline
{\small \textbf{Figure 2.} The entropy and specific heat of $k=0,-1$
solutions are represented as a function of $N$ for bolt solutions in six
dimensions. }
\\
\\

In Figure 2, we
ploted the entropy and specific heat per unit volume as function of $N$ for $%
\ell =1,|\sigma|=1.5$ for topological bolt solutions in six dimensions.  In
the lower branch case, the relative sign of $(S,C)$ alternates between
spacetime dimensionalities of $4m$ and $(4m+2)$.

Also, for all studied cases, the entropy is not proportional to the area,
despite the absence of Misner strings.


\section{Properties of the boundary metric}

Typically the boundary of this kind of spacetime will be an asymptotic
surface at some large radius $r$. However the metric restricted to the
boundary $\gamma _{ij}$ diverges due to an infinite conformal factor $%
r^{2}/\ell ^{2}$, and so the metric upon which the dual field theory resides
is usually defined using the rescaling 
\begin{eqnarray}  \label{resc}
h_{ab}=\lim_{r\rightarrow \infty }\frac{\ell ^{2}}{r^{2}}\gamma _{ab}.
\end{eqnarray}%
Corresponding to the boundary metric $h_{ab}$, the stress-energy tensor $%
<\tau _{ab}>$ for the dual theory can be calculated using the following
relation \cite{Myers:1999qn} 
\begin{eqnarray}  \label{r1}
\sqrt{-h}h^{ab}<\tau _{bc}>=\lim_{r\rightarrow \infty }\sqrt{-\gamma }\gamma
^{ab}T_{bc}.
\end{eqnarray}

Employing the prescription (\ref{resc}) we find the general line element of
the dual field theory for a Lorentzian $(D+1)$-dimensional TNAdS solution 
\begin{eqnarray}  \label{bound-gen}
ds^{2}=\ell ^{2}(d\theta _{i}^{2}+f_{k}^{2}(\theta _{i})d\varphi _{i}^{2})-%
\Big(4nf_{k}^{2}(\frac{\theta _{i}}{2})d\varphi _{i}+dt\Big)^{2},
\end{eqnarray}%
where again the index $i$ goes from $1$ to $p$ and $f_{k}(\theta )$ is given
by eq.~(\ref{f}). Here it is convenient to use the new parameters $m=1/\ell
,~\Omega =n/\ell ^{2}$ and to define new ``radial'' coordinates $\rho
_{i}=\ell \theta _{i}$. With these conventions, the $D-$dimensional line
element (\ref{bound-gen}) becomes 
\begin{eqnarray}  \label{Godel-gen}
ds^{2}=d\rho _{i}^{2}+\frac{f_{k}^{2}(m\rho _{i})}{m^{2}}d\varphi _{i}^{2}-%
\Big(\frac{4\Omega }{m^{2}}f_{k}^{2}(\frac{m\rho _{i}}{2})d\varphi _{i}+dt%
\Big)^{2}.
\end{eqnarray}%
%
%
%
%
%
%
%
%
%

\subsection{The three dimensional case}

We start again with a discussion of the well-known three dimensional case $%
(p=1)$. For $k=-1$ the corresponding line element is given by 
\begin{eqnarray}  \label{godel-3d-1}
ds^{2}=d\rho ^{2}+\frac{\sinh ^{2}(m\rho )}{m^{2}}d\varphi ^{2}-\Big(\frac{%
4\Omega }{m^{2}}\sinh ^{2}(\frac{m\rho }{2})d\varphi +dt\Big)^{2},
\end{eqnarray}%
which is the standard form used in the literature of the nontrivial
(2+1)-dimensional part of a hyperbolic homogeneous G\"{o}del-type universe $%
\mathcal{G}_{3}$ \cite{reboucas}, which can also be viewed as a squashed AdS$%
_{3}$ solution \cite{Rooman:1998xf}. The four$-$dimensional metric is the
direct Riemannian sum $dz^{2}+ds^{2}$ of a flat factor and the $(2+1)-$dimensional metric (\ref{godel-3d-1}). However, the three dimensional
metric part contains all the interesting effects exhibited by a
four$-$dimensional G\"{o}del-type spacetime. The famous G\"{o}del rotating
universe \cite{Godel:1949ga} corresponds to the case $m^{2}=2\Omega
^{2}~(\ell ^{2}=2n^{2})$. The value $m^{2}=4\Omega ^{2}~(\ell ^{2}=4n^{2})$
is AdS$_{3}$ spacetime written in rotating coordinates; the four dimensional
generalization is known as the Rebou\c{c}as-Tiomno space-time \cite%
{Reboucas:wa}.

The boundary spacetime of a $k=0$ topologically-Taub-NUT-AdS solution
corresponds to the nontrivial $(2+1)$-dimensional part of the
Som-Raychaudhuri spacetime \cite{Som} 
\begin{eqnarray}  \label{godel-3d-0}
ds^2=d\rho^2+\rho^2 d\varphi^2-\Big( \Omega \rho^2 d\varphi+dt\Big)^2,
\end{eqnarray}
which also has some relevance in a string theory context \cite%
{Horowitz:1994rf}.

According to the terminology in ref.~\cite{reboucas}, there are also
circular $(k=1)$ metrics 
\begin{eqnarray}  \label{godel-3d+1}
ds^2=d\rho^2+\frac{\sin^2(m \rho)}{m^2}d\varphi^2-\Big(\frac{4\Omega}{m^2}
\sin^2(\frac{m \rho}{2})d\varphi+dt\Big)^2,
\end{eqnarray}
corresponding to the boundary of a $k=1$ four dimensional Taub-NUT-AdS
solution. Their properties are extensively discussed in ref.~\cite{reboucas}%
. These metrics can formally be obtained by taking $m\to im$ in eq.~(\ref%
{godel-3d+1}), while the line element (\ref{godel-3d-0}) corresponds to a
limit $m\to 0$.

A well known property of the line elements (\ref{godel-3d-1})-(\ref%
{godel-3d+1}) is the occurrence of CTCs for a range of the parameters ($%
m,\Omega $) \cite{reboucas}. The violation of causality is made possible
essentially because the metric coefficient $g_{\varphi \varphi }$ assumes
negative values for some range of values of $\rho $, an effect induced by
the nondiagonal metric term associated with rotation. This is quite
different from creating causal anomalies in flat space by identifying the
time coordinate or from the standard AdS causal problems. Furthermore,
because these spacetimes are homogeneous, there are CTCs through every event
(hence the causality violation is not localized to some small region). Also, a
study of geodesics has shown that these spacetimes are geodesically complete
(and thus singularity free) \cite{Chandrasekhar,Calvao:1990yv}.

For $k=1,0$ causality is violated for every ($m,\Omega $). In the hyperbolic
case $k=-1$, causality violation on the boundary metric (\ref{godel-3d-1})
will appear only for $m^{2}<4\Omega ^{2}$, which separates globally
hyperbolic spacetimes from causality violating ones.

For $k=0,-1$, a natural choice for the range of the coordinates covering the
entire manifold is $-\infty <t<\infty $, $0\leq \rho <\infty $, $0<\varphi
\leq 2\pi $. For $k=1$, it is natural to take $0\leq \rho \leq \pi m$, $%
0<\varphi \leq 2\pi $ while, similar to the bulk case, the absence of Dirac
string singularities imposes a time periodicity $\beta =8\pi \Omega /m^{2}$
(which is just the bulk periodicity).


\subsection{Five dimensional G\"odel-type solutions}

Five dimensional line-elements of the form (\ref{Godel-gen}) have recently
been of considerable interest, after the discovery \cite{Gauntlett:2002nw}
of a five dimensional G\"{o}del-type supersymmetric solutions with very
similar properties to $k=0$ solution in three dimensions. The line-element
of this solution reads 
\begin{eqnarray}  \label{5dk=0}
ds^{2}=d\rho _{1}^{2}+\rho _{1}^{2}d\varphi _{1}^{2}+d\rho _{2}^{2}+\rho
_{2}^{2}d\varphi _{2}^{2}-\Big(\Omega (\rho _{1}^{2}d\varphi _{1}+\rho
_{2}^{2}d\varphi _{2})+dt\Big)^{2},
\end{eqnarray}%
corresponding to the case $p=2,k=0$ in eq.~(\ref{Godel-gen}). Here again,
although $\partial /\partial t$ is an everywhere timelike Killing vector,
the coordinate $t$ fails to be a global time function. As explicitly proven
in ref.~\cite{Boyda:2002ba}, this geometry is geodesically complete and
also\ contains nontrivial CTCs.

Similar solutions can be written for $k=\pm 1$. However, although possessing
the same symmetries, these are generally not solutions of the field
equations of the minimal supergravity, and new matter fields should be
introduced. In this context, note also the existence of a different
construction of the five-dimensional G\"{o}del-type line element, within a
different metric ansatz \cite{Reboucas:1998qg}.

The hyperbolic case $k=-1$ of $\mathcal{G}_{5}$ is also of special interest;
its line element is of the form 
\begin{eqnarray}
ds^{2} &=&d\rho _{1}^{2}+\frac{1}{m^{2}}\sinh ^{2}(m\rho _{1})d\varphi
_{1}^{2}+d\rho _{2}^{2}+\frac{1}{m^{2}}\sinh ^{2}(m\rho _{2})d\varphi
_{2}^{2}  \nonumber  \label{5dk=-1} \\
&&-\Big(\frac{4\Omega }{m^{2}}\sinh ^{2}(\frac{m\rho _{1}}{2})d\varphi _{1}+%
\frac{4\Omega }{m^{2}}\sinh ^{2}(\frac{m\rho _{2}}{2})d\varphi _{2}+dt\Big)%
^{2}.
\end{eqnarray}%
The solution with $m^{2}=6\Omega ^{2}$ solves the Einstein equations with a
negative cosmological constant $\Lambda =-6\Omega ^{2}$. This solution of
minimal gauged supergravity admits only one Killing spinor \cite%
{Gauntlett:2003fk}. It is therefore not maximally supersymmetric (it is not 
AdS$_{5}$) \cite{Britto-Pacumio:1999sn}. Furthermore for $%
m^{2}<4\Omega ^{2}$ the Killing vectors $\partial /\partial \varphi _{i}$
become timelike for large enough values of $\rho _{i}$, while $f(x^{i})=t$
is a global time coordinate only for $m^{2}>8\Omega ^{2}$.


\subsection{Properties of the general case}

The line element (\ref{Godel-gen}) provides a natural generalization of the
well-known $\mathcal{G}_{3}$ line element, presenting many similar
properties. For every set $(\rho _{i},\varphi _{i})$ and value of $k$, it is
natural to take a coordinate range very similar to the $\mathcal{G}_{3}$
case. Also, all $k=1,0$ metrics have nontrivial CTCs passing through every
spacetime point (though, as for the G\"{o}del universe, these are not
geodesics). Therefore these manifolds cannot be foliated by everywhere
spacelike surfaces and the classical Cauchy problem is ill posed.

Again, the case $k=-1$ is rather special, since for $m^{2}>4p\Omega ^{2}$
the time coordinate $t$ is a global time function and the spacetimes are
globally hyperbolic. It can also be proven that $\mathcal{G}_{2p+1}$ is a
homogeneous spacetime for every choice of $k$.

These singularity-free spacetimes have a Ricci scalar $R=2p(\Omega
^{2}-m^{2})$. Let us introduce an orthonormal basis with $ds^{2}=\eta
_{ij}\vartheta ^{i}\vartheta ^{j}$ (where $\vartheta ^{\rho ^{i}}=d\rho
^{i},~\vartheta ^{\varphi ^{i}}=f_{k}(\rho ^{i})d\varphi ^{i},~\vartheta
^{2p+1}=dt+\frac{4\Omega }{m^{2}}f_{k}^{2}(m\rho _{i}/2)d\varphi _{i}$) with $%
i,j=1,\dots p$. In this frame, we find the nonvanishing components of the Einstein tensor 
\[
E_{\rho _{i}}^{\rho _{i}}=E_{\varphi _{i}}^{\varphi _{i}}=\Omega
^{2}(2-p)+km^{2}(1-p),~~E_{t}^{t}=p(km^{2}-3\Omega ^{2}). 
\]%
Thus the energy-momentum tensor for these spacetimes is that of a perfect
fluid $T_{a}^{b}=(P+\epsilon )u_{a}u^{b}+P\delta _{a}^{b}$, with pressure $%
P=\Omega ^{2}(2-p)-km^{2}(1-p)$ and energy density $\epsilon =p(3\Omega
^{2}-km^{2})$ while $u^{a}=\delta _{t}^{a}$ is the velocity of a fundamental
class of observers. For $km^{2}=2\Omega ^{2}(p+1)$ this corresponds to a
negative cosmological constant $\Lambda =-p(2p-1)\Omega ^{2}$. It would be
interesting the find a matter content compatible with these geometries in a
field theory context.

The high degree of symmetry of the line element (\ref{Godel-gen}) simplifies
the analysis of some basic physical processes. For example the geodesic
equations can easily be solved in the general case. The geodesic equations
have the set of independent first-integrals 
\begin{eqnarray}
\dot{t} &=&E-4\Omega \frac{f_{k}^{2}(m\rho _{i}/2)}{f_{k}^{2}(m\rho _{i})}%
\Big(L_{i}+\frac{4\Omega E}{m^{2}}f_{k}^{2}(m\rho _{i}/2)\Big),  \nonumber \\
\dot{\varphi _{i}} &=&\frac{m^{2}}{f_{k}^{2}(m\rho _{i})}\Big(L_{i}+\frac{%
4\Omega E}{m^{2}}f_{k}^{2}(m\rho _{i}/2)\Big),  \nonumber
\end{eqnarray}%
where $E,~L_{i}$ are integration constants implied by the existence of the
Killing vectors $\partial /\partial t$ and $\partial /\partial \varphi _{i}$%
. A superposed dot stands for as derivative with respect to the parameter $%
\tau $, which is an affine parameter along the geodesics; for timelike
geodesics, $\tau $ is the proper time. Thus we find a set of $p$ equations
on the form 
\begin{eqnarray}  \label{eqgen}
\dot{\rho _{i}}^{2}=\alpha _{i}^{2}-\frac{m^{2}}{f_{k}^{2}(m\rho _{i})}\Big(%
L_{i}+\frac{4\Omega E}{m^{2}}f_{k}^{2}(m\rho _{i}/2)\Big),
\end{eqnarray}%
where $\alpha _{i}$ are constants satisfying the constraint $\sum_{i}\alpha
_{i}^{2}=E^{2}-\varepsilon $ ($\varepsilon =1$ or $0$ for timelike or null
geodesics respectively). This type of equation is known from the study of
geodesic motion in $\mathcal{G}_{3}$ spacetimes \cite{Calvao:1990yv}. All
the properties derived there apply directly to the general case. For example
the geodesic motion is always bounded in the coordinates $\rho _{i}$, except
for those free particles in the hyperbolic family of spacetimes which have $%
\alpha _{i}^{2}-4\Omega ^{2}E^{2}/m^{2}>0$ (this is possible only for values
of $m^{2}>4\Omega ^{2}$). The equation (\ref{eqgen}) can be solved by taking $%
x_{i}=f_{k}^{2}(m\rho _{i}/2) $, which leads to the solution 
\begin{eqnarray}
f_{k}^{2}(\frac{m\rho _{i}}{2})=\frac{1}{2( 4\Omega ^{2}E^{2}/m^{2}+k\alpha
_{i}^{2})}\Big(\alpha _{i}^{2}-2\Omega EL_{i}-\alpha _{i}\sqrt{\alpha
_{i}^{2}-4\Omega L_{i}E-km^{2}L_{i}^{2}}  \nonumber \\
\times \cos m\sqrt{k\alpha _{i}^{2}+\frac{4\Omega ^{2}E^{2}}{m^{2}}}(\tau
-\tau _{0})\Big),  \nonumber
\end{eqnarray}%
which is valid for all $k=1,0$ solutions and the $k=-1$ solutions with $%
\alpha _{i}^{2}-4\Omega ^{2}E^{2}/m^{2}<0$. The unbounded hyperbolic motion
is given by the equation 
\begin{eqnarray}
\sinh ^{2}(\frac{m\rho _{i}}{2})=\frac{1}{2( 4\Omega ^{2}E^{2}/m^{2}-\alpha
_{i}^{2})}\Big(\alpha _{i}^{2}-2\Omega EL_{i}-\alpha _{i}\sqrt{\alpha
_{i}^{2}-4\Omega L_{i}E+m^{2}L_{i}^{2}}  \nonumber \\
\times\cosh m\sqrt{\alpha _{i}^{2}-\frac{4\Omega ^{2}E^{2}}{m^{2}}}(\tau
-\tau _{0})\Big).  \nonumber
\end{eqnarray}%
Thus, it follows that, similar to $\mathcal{G}_{3}$, for $k=-1$ metrics with 
$m^{2}<4\Omega ^{2}$ each geodesic observer has an optical horizon, $i.e.$ a
maximum spatial distance in a $(\rho _{i},\varphi _{i})$ plane beyond which
it is not possible to receive light signals. 

The wave equation for a scalar, a Dirac or an electromagnetic field can also
easily be solved in term of hypergeometric functions for any $(p,k)$.


\section{Discussion}

In this paper we have considered higher dimensional generalizations of the
known four dimensional topological nut and bolt metrics with negative
cosmological constant. In these solutions, angular spheres with coordinates $%
(\theta ,\varphi )$ are replaced by planes and hyperboloids.

These spacetimes with nut charges present interesting examples of geometries
with CTCs. Unlike the $k=1$ case, the CTCs arise even in the $k=0,-1$ cases 
were no temporal identifications are made. Analysis of the
thermodynamics of such solutions is problematic in a Lorentzian context.
However it is fairly straightforward to conduct such an analysis on their
Euclidean sections. In particular, gravitational entropy (defined via the
relation (\ref{GibbsDuhem})) will be present whenever it is not possible to
foliate a given spacetime in the Euclidean regime\footnote{%
Here we refer to a Euclidean geometry with non trivial topology. The
topology of a Lorentzian spacetime can change with time, only if there is
some pathology, such as a singularity or closed time like curves.} by a
family of surfaces of constant time \cite{Hawking:1998ct}.

If the topology of the $\left( D+1\right) -$dimensional Euclidean section
is such that there are fixed points in the U(1) isometry group generated by the (Euclidean)
timelike Killing vector $\xi =\partial /\partial \tau $ then it will not be
possible to everywhere foliate the spacetime with surface of constant $\tau $%
. In general conical singularities will appear unless $\tau $ is assigned
the requisite periodicity, yielding the usual relationship between area and
entropy provided the co-dimension of the fixed-point set is $\left(
D-1\right) $. Black hole horizons are bolts but not all bolts are horizons. That is 
not every Euclidean solution with a bolt has an analytic continuation to a
Lorentzian section that is a black hole. However in the presence of NUT\ charge 
the usual (and presumed universal) relationship between black hole area and entropy 
is generalized.

Remarkably enough, the AdS/CFT counterterm approach can be employed to
compute conserved quantities and gravitational entropy even in these more
general cases \cite{Clarkson:2002uj}.

Our investigation of the thermodynamics of the topologically TNAdS solutions
has yielded a number of unexpected results. The properties of the four
dimensional solutions are somewhat generic. For any dimension, the planar
nuts represent extremal backgrounds, with zero temperature. Furthermore, for
negative $\Lambda $ there are no hyperbolic nuts. In addition, we found that
the first law of thermodynamics removes the ambiguity in the choice of bolt
radius $r_{b}$ and fixes its value as a function of ($N,\ell )$, rather
similar to the known $k=1$ spherically symmetric case. This in turn yields
the result of eq. (\ref{temp}) relating $\beta $ and the NUT charge $N$, the
key distinction being that for the $k=1$ case the parameter $\sigma $ is an
integer, whereas for the $k=0,-1$ cases it is arbitrary.

The entropy derived within the counterterm approach in all cases $k=1,0,-1$
fails to satisfy the usual entropy-area relationship, despite the absence of
Misner string singularities in the latter two cases. This is perhaps the
most surprising result of this paper: it seems that the entropy/area
relation is always violated in the presence of a NUT charge. We also found
the rather unusual situation that, although the $k=0,-1$ solutions possess 
nonzero angular momenta, these charges do not affect the thermodynamics of
the system.

It is a well-known fact that a phase transition occurs for hot AdS. That is,
at a critical temperature, the thermal radiation in AdS will collapse
forming a black hole. Equivalently, the low temperature black holes are not
stable and global AdS spacetime is then the preferred state. On the other
hand, high temperature $large$ black holes are stable\footnote{%
This was observed by Hawking and Page \cite{hawking1}. Euclidean sections of
black hole metrics are periodic in the imaginary time and so, they represent
black holes in equilibrium with thermal radiation. However, the thermal
radiation in asymptotically flat spaces (all the way to infinity) will have
an energy density that will curve the spacetime --- make it an expanding or
collapsing Friedmann universe. Then, a static situation can be obtained if
one considers a black hole in a finite sized box. On the other hand, it is
well-known that Schwarzschild black holes have negative specific heat ---
they absorb faster than they radiate and the equilibrium is unstable.
Working in AdS spacetime, one solves both problems: the gravitational
potential in AdS space acts like a confining box and black holes larger than
the radius of curvature of AdS have positive specific heat (and are
presumably stable).}--- they do not decay to global AdS spacetime. From the
gauge theory viewpoint, this is a confinement/deconfinement phase transition %
\cite{witten}.

The thermodynamics of spherical Taub-Bolt-AdS resembles the
Schwarzschild-AdS case. That is, there is a phase transition between $k=1$
Taub-Bolt-AdS and Taub-NUT-AdS. The idea is that one must consider all
possible spacetime solutions with the appropriate boundary conditions ---
possible saddle points in the Euclidean path integral over gravity in the
bulk (Taub-Bolt-AdS and Taub-NUT-AdS in our case). The Taub-NUT-AdS metric
dominates the path integral at low temperatures, whereas the the
Taub-Bolt-AdS metric dominates at high temperatures.

However, unlike the spherical ($k=1$) Taub-Bolt-AdS spaces, the planar ($k=0$%
) Taub-Bolt-AdS spaces have the property that they do not exhibit a phase
transition at finite temperature.\footnote{%
As emphasized in ref.~\cite{buchel}, even if classical (Euclidean) $AdS_{5}$
foliations by $R^{4},S^{4}$ and $H_{4}$ are related by coordinate
transformations, the corresponding gauge theories are physically
inequivalent --- there is a phase transition for $k=1$ but not for $k=0$ (in
this case there is a phase transition if one considers the AdS soliton
instead of AdS \cite{Surya:2001vj}). Different space-like foliations of the
background geometry lead to different definitions of time (and the
Hamiltonian) of a quantum system. The classical supergravity background
(large $N$ limit and large 't Hooft coupling) is equivalent to the full
quantum gauge theory on the corresponding slices.}

One determines if there is a phase transition by comparing the Euclidean
action of the Taub-Nut-AdS and Taub-Bolt-AdS metrics. For $k=0$, we obtain 
\[
I_{b}-I_{n}=-\frac{VD\beta }{16\pi G\ell ^{2}}\int_{N}^{r_{b}}ds\frac{%
(s^{2}+N^{2})}{2s^{2}}(s^{2}-N^{2})^{(D-1)/2}, 
\]
which is a negative quantity for any choice of $(N,\ell )$.

We also considered the basic properties of the boundary metric of the
Taub-NUT-AdS Lorentzian solutions. This led to a natural $\mathcal{G}_{n}$
generalization to higher dimensions of the famous $\mathcal{G}_{3}$ G\"{o}%
del-type geometry. These spacetimes also contain CTCs for a large range of
the parameters, and are interesting especially in connection with the
recently discovered five dimensional supersymmetric G\"{o}del solution.

Our computations have been carried out in the Euclidean
section. At this point we remark that, on the CFT side, the analytic
continuation (\ref{cont}) corresponds on the boundary to the Euclidean
approach to field quantization in the presence of CTCs, originally proposed
by Hawking \cite{Hawking:1995zi} 
\[
t\rightarrow i\tau ,~~\Omega \rightarrow ia. 
\]%
In this approach we analytically continue the Lorentzian spacetime to get a
metric with a real Euclidean section. On this section, all the field
operators commute and the Green functions are well defined. Explicit values
of field operators can be defined by analytic continuation from the
Euclidean spacetime. Here we remark that in the presence of off-diagonal
terms of the metric tensor, Wick rotation is more problematic and generally
involves the analytic continuation of more parameters than the time
coordinate ($e.g.$, the angular momentum $J$ for a Kerr-Newmann black hole).
For a G\"{o}del-type spacetime, we have to supplement the analytic
continuation $t\rightarrow i\tau $ with a continuation in the rotation
parameter $\Omega $. However, in contrast to the asymptotic meaning of
rotation (as opposed to the angular momentum $J$) for a Kerr-Newmann black
hole, in this case the rotation has a well defined local character.

By analytic continuation, we obtain a Euclidean metric of the general form
\begin{eqnarray}  \label{gen-eucl-1}
ds^{2}=d\rho _{i}^{2}+\frac{f_{k}^{2}(m\rho _{i})}{m^{2}}d\varphi _{i}^{2}+%
\Big(\frac{4a}{m^{2}}f_{k}^{2}(\frac{m\rho _{i}}{2})d\varphi _{i}+d\tau \Big)%
^{2},
\end{eqnarray}
which is also (after a suitable rescaling) the boundary of the $(D+1)$-dimensional 
Euclidean TNAdS spacetime.

It would be desirable to compute some quantities in a Euclideanized G\"{o}%
del-type background and to compare the results with the bulk predictions, in
order to see how much the supergravity ``knows'' about the boundary CFT. Since 
in the cases of interest in this paper the dual conformal field theories are 
poorly known, it is impossible to discuss them in detail. The best one can 
do at this point is to compare the thermodynamics of a scalar field 
(toy example) on a Euclideanized G\"{o}del-type background with the results 
obtained on the gravity side.

Given the high degree of symmetry of the boundary metrics, the
well-known formalism of zeta function regularization \cite{Hawking:1976ja}
and its recently proposed generalization (the ``local zeta function
approach''  \cite{Moretti:1997qn}) are especially well suited. This gives
rigorous results for the effective action, the vacuum expectation value of
the field fluctuations and $<\tau _{b}^{a}>$ for a quantum field propagating
in a zero- or finite temperature Euclideanized G\"{o}del-type background.
Unfortunately, the computations are quite complicated since the involved
series do not occur in the literature. For $k=1$, this procedure implies a
generalization of the methods presented in ref.~\cite{Dowker:1998pi} for $D=3
$. To our knowledge, the cases $k=0,-1$ have not been discussed in the
literature.

A computation of the effective action for a scalar field propagating in a
zero-temperature Euclidean $\mathcal{G}_{2p+1}$ background with $k=0$
(including the five-dimensional supersymmetric case \cite{Gauntlett:2002nw})
is presented in Appendix B. A direct application of the relations (\ref%
{defIeff}), (\ref{zetagen}) leads to an effective action for a massless,
minimally coupled scalar field in three and five dimensions 
\begin{eqnarray}
I_{eff}^{D=3} &=&3\beta Vl^{2}a\zeta _{R}^{\prime }(-2)\simeq
-0.0913454\beta Vl^{2}a,  \nonumber \\
I_{eff}^{D=5} &=&-\beta Vl^{4}a(\zeta _{R}^{\prime }(-2)+2\zeta _{R}^{\prime
}(-4))\simeq 0.00482694\beta Vl^{4}a,  \nonumber
\end{eqnarray}%
the general expression of the effective action being on the form 
\beqa
\label{Ieff-gen}
I_{eff}\sim \beta V\ell ^{D-1}a,
\eeqa
where $\zeta _{R}(s)$ is the usual Riemann zeta function. Here $\beta $ is
some (large-) periodicity of the coordinate $\tau $ necessary to ensure the
orthonormality of the scalar field eigenfunctions. Corresponding expressions
for the massive, nonminimally coupled case can also be written by using (\ref%
{defIeff}, \ref{zetagen-c}). These results would make the connection with
the $(D+1)$-dimensional $k=0$ NUT case, since from the AdS/CFT
correspondence we expect $I_{eff}$ for a massless, conformally coupled
scalar field to have a similar form to the bulk action (\ref{gen1-3}) (we
recall that $a=N/\ell ^{2}$). We note here that the bulk actions contain 
a $(N/\ell )^{D-1}$ scaling factor, as compared to the above
$I_{eff}$. We also found the same signs for the actions.

Although the extrapolation of the zeta function expression
(\ref{zetagen-c}) to the massless, conformally coupled case is not
straightforward, since the corresponding scalar field spectrum
contains negative eigenvalues, we do not expect to find there a
different result. Based on the results in Appendix B, we predict in
this case a dependence of the effective action on the parameters
$(a,\ell )$ on the form (\ref{Ieff-gen}), and the same different
scaling factors $(N/\ell )^{D-1}$. Thus, the overall normalization
of the action seems to require a knowledge of physics at string
scale curvatures.

The calculations are done in different regimes: the weakly coupled regime for
the CFT (open string) and the strongly coupled regime for gravity (closed
string). However, the results appear to be remarkably close. An explanation 
could be the fact that there are no phase transitions as a function of
temperature. Then, the theory would appear to be free from drastic changes 
of degrees of freedom as the coupling is increased.

We close with some comments on the dual Lorentzian conformal field theory.
Over the last decade, the issue of causality violation has been extensively
discussed in the literature, with special attention being paid to field
quantization on a spacetime background that contains CTCs (see ref.~\cite%
{Visser:2002ua} for a recent review). It was found that such spacetimes are
usually unstable with respect to quantum fluctuations. However, most of the
papers deal with confined causality violating spacetimes, $i.e.$, the CTCs
are confined within some region and there exists at least one region free of
them. Regions with CTCs are separated from those without CTCs by Cauchy
horizons. Obviously, this is not the case for the G\"{o}del spacetime where
causality violation is not a result of the evolution of certain initial
data, but rather has existed ``forever''.

Field quantization in a G\"{o}del-type spacetime has received scant
attention in the literature. The difficulties in the standard field
quantization in G\"{o}del universe have been explicitly pointed out by Leahy
in ref.~\cite{Leahy:1982dj} and consist mainly of the absence of a complete
Cauchy surface and of the incompleteness of the mode solutions to the field
equations. Despite some attempts, the meaning of a quantum field theory in
this background is still unclear.

This issue is interesting especially in connection with the chronology
protection conjecture \cite{Hawking:1991nk}. A repeated number of attempts
have suggested that the back reaction of the (divergent) energy momentum
tensor of a test field on the spacetime geometry, via the semi-classical
Einstein equations, could furnish a possible mechanism for the enforcement
of chronology protection. The status of this conjecture in a G\"{o}del-type
background is still unsettled.

We argue here that a new perspective on this issue may come from a rather
unexpected direction: the conjectured AdS/CFT correspondence. Assuming the
validity of this conjecture, the NUT charged AdS solutions provide an
opportunity for the study of quantum field theories on G\"{o}del-type
rotating spacetimes, which is clearly an interesting subject. Implicitly, a
consistent formulation of the field quantization in a G\"{o}del-type
background should be possible (although physical intuition tells us that
such a theory should present some pathological properties). According to the
standard approach, this is possible only for causally well behaved metrics
(corresponding for $\mathcal{G}_{3}$ case to $k=-1,~m^{2}\geq 4\Omega ^{2}$).

Of course, similar to many other situations we do not know here the dual CFT
theory. However, we can use the AdS/CFT ``dictionary'' to predict
qualitative features of a quantum field theory in these backgrounds. 
For example, as implied by this correspondence, the boundary stress tensor
in the AdS-NUT space should match the expectation value of the dual CFT. For
the four-dimensional Lorentzian line element (\ref{metric-4D}), the
prescription (\ref{r1}) gives for the stress tensor of the dual theory in a $%
\mathcal{G}_{3}$ background a simple expression 
\[
<\tau _{b}^{a}>=\frac{Mm^{2}}{8\pi G}[3u^{a}u_{b}+\delta _{b}^{a}],
\]%
where $u^{a}=\delta _{t}^{a}$. This tensor is finite, covariantly conserved
and manifestly traceless (and is the standard form for the stress tensor of
a $(2+1)$-dimensional CFT). Applying the same procedure to a six-dimensional
TNAdS metric yields a somewhat similar result for the stress tensor of the
dual theory in the five-dimensional supersymmetric $\mathcal{G}_{5}$
background 
\[
<\tau _{b}^{a}>=\frac{Mm^{4}}{8\pi G}[5u^{a}u_{b}+\delta _{b}^{a}],
\]%
which is also finite in every spacetime point, even when the CTCs appear.
Thus CTCs do not necessarily mean that the energy-momentum tensor must
diverge.\footnote{A similar result has recently been obtained \cite%
{Gauntlett:2004cm} for a one-parameter family of supersymmetric solutions of
type IIB supergravity that includes AdS$_{5}\times S^{5}$ ; it also has
causality-violating configurations.} Hence we cannot use this
argument as support for the proposal that  causality violation occurring in
a G\"{o}del-type spacetime might be removed by quantum effects. This situation 
is different from the
confined causality violating spacetimes, where the energy momentum tensor of
a quantum field diverges at the Cauchy horizon --- unlike G\"{o}del-type
spacetimes, there are geodesic CTCs which dominate quantum amplitudes in a
semiclassical picture.

The finiteness of the stress tensor of a CFT on G\"{o}del background and of
the effective action on the Euclidean/Lorentzian section (or equivalently
the entropy) reveals an important fact: in this case, chronology protection
cannot be settled at the level of semiclassical quantum gravity ($i.e.$ the
gravity background is classical but the matter fields are quantized).

Since string theory is a proposed fundamental theory of quantum gravity, one
may expect that stringy corrections to the low energy effective theory resolve
causality violations. The low energy effective equations of string theory
have as solutions an abundance of such causality violating geometries. Many
attempts have been made to understand or to excise the regions with CTCs by
holographic screens \cite{Boyda:2002ba,Brecher:2003rv} or by probing the
geometry\footnote{%
This resembles the enhan\c{c}on mechanism (see, $e.g.$, refs.~\cite{enhancon}%
).} with appropriate probes \cite{probes}. Such a stringy mechanism for the
Goedel-like universe was proposed by Drukker et al. in ref.~\cite{nadav}
(further refined in the subsequent work of ref.~\cite{nadav2}). Using
supertubes \cite{supertubes} as probes in a IIA G\"{o}del universe, they
show that there will appear instabilities in the CTC region (despite the
fact that supertubes preserve the same supersymmetries as the G\"{o}del-like
universe itself), concluding that this geometry is not a good string theory
background.


\section*{Acknowledgements}

The authors would like to thank Alex Buchel, Igor Drozdov, Nadav Drukker,
David Mateos, David Page and Cristian Stelea for valuable discussions. DA 
also thanks Rob Myers for useful conversations. DA gratefully acknowledges 
the ongoing hospitality of the University of Waterloo's Department of Physics 
during this research.

DA is supported by a Dow Hickson Fellowship. The work of RBM was supported
in part by the Natural Sciences and Engineering Research Council of Canada.
The work of ER was supported by Graduiertenkolleg of the Deutsche
Forschungsgemeinschaft (DFG): Nichtlineare Differentialgleichungen;
Modellierung, Theorie, Numerik, Visualisierung and Enterprise--Ireland Basic
Science Research Project SC/2003/390 of Enterprise-Ireland.

\appendix

\section{Misner string analysis for general $k$}

We consider here the question of Misner string singularities for
compactified  Taub-NUT-AdS spacetimes. Althought we set
$D=3$, the discussion below is straightforwardly generalized to
other values of $D$ --- also, it is not dependent upon the specific 
form of $F(r)$ in eq.~(\ref{F}).

Following the analysis ~of Misner \cite{Misner}, we work with the
following set of basis vectors 
\begin{eqnarray}
\nonumber
\omega ^{0} &=&\sqrt{F(r)}\left( dt+4nf_{k}^{2}(\frac{\theta }{2})d\varphi
\right) ,   \label{4} 
\\
\omega ^{1} &=&\frac{dr}{\sqrt{F(r)}} ,  \label{5}
\\
\nonumber
\omega ^{2} &=&\sqrt{r^{2}+n^{2}}d\theta ,   \label{6}
\\
\nonumber
\omega ^{3} &=&\sqrt{r^{2}+n^{2}}f_{k}(\theta )d\varphi ,  \label{7}
\end{eqnarray}%
in which case the metric (\ref{metric-4D}) becomes
\begin{eqnarray}
\nonumber
ds^{2}=-\left( \omega ^{0}\right) ^{2}+\left( \omega ^{1}\right) ^{2}+\left(
\omega ^{2}\right) ^{2}+\left( \omega ^{3}\right) ^{2}.  \label{8}
\end{eqnarray}%
The curvature scalars are finite and so there is no curvature singularity.
Inverting eqs. (\ref{5}) yields
\begin{equation}
dt=\frac{1}{\sqrt{F}}\omega ^{0}-\frac{4nf_{k}^{2}(\frac{\theta }{2})}{\sqrt{%
r^{2}+n^{2}}f_{k}(\theta )}\omega ^{3}.  \label{10}
\end{equation}%
Thus, we obtain 
\begin{equation}
-\left( \nabla t\right) ^{2}=\frac{1}{F}-\frac{\left( 4n\right) ^{2}}{\left(
r^{2}+n^{2}\right) }\left[ \frac{f_{k}^{2}(\frac{\theta }{2})}{f_{k}(\theta )%
}\right] ^{2},  \label{11}
\end{equation}%
which will become infinite whenever $\frac{f_{k}^{2}(\frac{\theta }{2})}{%
f_{k}(\theta )}$ diverges. Explicitly we have 
\begin{equation}
\frac{f_{k}^{2}(\frac{\theta }{2})}{f_{k}(\theta )}=\left\{ 
\begin{array}{ll}
\frac{1}{4}\tan ^{2}\left( \frac{\theta }{2}\right) , & \mathrm{for}\ \ k=1
\\ 
\frac{\theta ^{2}}{16}, & \mathrm{for}\ \ k=0 \\ 
\frac{1}{4}\tanh ^{2}\left( \frac{\theta }{2}\right) , & \mathrm{for}\ \
k=-1.%
\end{array}%
\right.   \label{12}
\end{equation}%
and so one can easy see that only for $k=1$ is there a singularity. If there
are identifications in the metric for the other two values of $k$, this will
yield $h\left( \theta ,\phi \right) =$constant, $i.e.$ $\theta $ will be
some function of $\phi $ at the identification boundaries. For example, a
torus can be described as a diamond in the plane with opposite sides
identified. We identify the boundary $y=1-x$ of the upper right of the
diamond with the boundary $y=-1-x$ on the lower left. This is the same as
saying that the curve $\theta \left( \cos \phi +\sin \phi \right) =1$ is
identified with the curve $\theta \left( \cos \phi +\sin \phi \right) =-1$
(writing $x=\theta \cos \phi $ and $y=\theta \sin \phi $). Nowhere does this
identification make $\left( \nabla t\right) ^{2}$ singular.

One can look at this in a different way by going to rectangular coordinates.
We use the relations 
\begin{equation}
r^{2\left| k\right| }\left( d\theta ^{2}+f_{k}^{2}(\theta )d\varphi
^{2}\right) =dx^{2}+dy^{2}+k\left( dz^{2}-dr^{2}\right)  \label{13}
\end{equation}%
where the coordinate transformation 
\begin{eqnarray}
\nonumber
x &=&r^{\left| k\right| }f_{k}(\theta )\cos \phi, \label{gigi}
\\
y &=&r^{\left| k\right| }f_{k}(\theta )\sin \phi,  \label{gigi1}
\\
\nonumber
z &=&r^{\left| k\right| }\sqrt{1+kf_{k}^{2}(\theta )}, \label{gigi3}
\end{eqnarray}%
has been employed. With this choice of variables the metric becomes 
\begin{equation}
ds^{2}=-\left( \omega ^{0}\right) ^{2}+\frac{dr^{2}}{F(r)}+\frac{r^{2}+n^{2}%
}{r^{2\left| k\right| }}\left( dx^{2}+dy^{2}+k\left( dz^{2}-dr^{2}\right)
\right)  \label{17}
\end{equation}%
for all three possible values of $k$. Since $x,y,z$ are all differentiable
functions, so is $r$. The only possible singularities in the metric for $%
F(r)\neq 0$ ($i.e.$ for sufficiently large $r$) are the singularities in $%
\omega ^{0}$, which has the form 
\begin{equation}
\omega ^{0}=\sqrt{F(r)}\left( dt+\frac{2^{\left| k\right| }n}{\left[ r\left(
z+r\right) \right] ^{\left| k\right| }}\left( xdy-ydx\right) \right)
\label{19}
\end{equation}%
and is singular at $z=-r$ only for $k=1$ --- for $k=0$ there are no
singularities. For $k=-1$, we obtain from eq.~(\ref{gigi1}) that $z=r\sqrt{%
1-f_{-1}^{2}(\theta )}\geq r$, and again there are no singularities. The
identifications are always in the $\left( x,y\right) $ plane for the $k=0,-1$
cases, and so do not render $\omega ^{0}$ singular.

Explicitly for each case we have --- for the sphere%
\begin{eqnarray}
x &=&r\sin \theta \cos \phi,   \nonumber \\
y &=&r\sin \theta \sin \phi,   \nonumber \\
z &=&r\cos \theta,   \nonumber \\
\omega ^{0} &=&\sqrt{F(r)}\left( dt+\frac{2n}{\left[ r\left( z+r\right) %
\right] }\left( xdy-ydx\right) \right),   \nonumber \\
ds^{2} &=&-\left( \omega ^{0}\right) ^{2}+\frac{dr^{2}}{F(r)}+\frac{%
r^{2}+n^{2}}{r^{2}}\left( dx^{2}+dy^{2}+\left( dz^{2}-dr^{2}\right) \right) ,
\nonumber
\end{eqnarray}
for the torus%
\begin{eqnarray}
x &=&\theta \cos \phi, \nonumber   \\
y &=&\theta \sin \phi,  \nonumber \\
\omega ^{0} &=&\sqrt{F(r)}\left( dt+n\left( xdy-ydx\right) \right), 
\nonumber \\
ds^{2} &=&-\left( \omega ^{0}\right) ^{2}+\frac{dr^{2}}{F(r)}+\left(
r^{2}+n^{2}\right) \left( dx^{2}+dy^{2}\right), \nonumber
\end{eqnarray}%
and for the hyperboloid%
\begin{eqnarray}
x &=&r\sinh \theta \cos \phi, \nonumber   \\
y &=&r\sinh \theta \sin \phi,   \nonumber \\
z &=&r\cosh \theta, \nonumber   \\
\omega ^{0} &=&\sqrt{F(r)}\left( dt+\frac{2n}{\left[ r\left( z+r\right) %
\right] }\left( xdy-ydx\right) \right), \nonumber \\
ds^{2} &=&-\left( \omega ^{0}\right) ^{2}+\frac{dr^{2}}{F(r)}+\frac{%
r^{2}+n^{2}}{r^{2}}\left( dx^{2}+dy^{2}-\left( dz^{2}-dr^{2}\right) \right). 
\nonumber
\end{eqnarray}

Therefore, the analysis above explicitly indicates an absence of
Misner-string singularities for $k=0,-1$.

\section{Scalar field in a $k=0$ Euclidean G\"odel-type background}

It would be desirable to compute some quantities in an Euclideanized
G\"odel-type background and to compare the results with the bulk
predictions. Although the details of the boundary CFT will depend on the
details of the bulk supergravity theory, the generic properties and possible
pathologies are expected to be independent of the precise features of the
theory. Therefore, similar to other situations we can restrict ourselves to
the simple case of a nonminimally coupled scalar field.

For a scalar field with an action 
\begin{eqnarray}  \label{s-action}
I[\phi]=- {\frac{1}{2}}\int \left(\nabla^\mu\phi\,\nabla_{\mu}\phi+M^2\phi^2
+\xi R\phi^2\right)\sqrt{g(x)}\,d^Dx,
\end{eqnarray}
(where $M$ is the scalar field mass and $\xi$ determines the coupling with
the scalar curvature $R$ of a $D-$dimensional G\"odel-type spacetime) the
zeta function approach implies the computation of the the eigenfunctions $%
\Psi_N$ and the eigenvalues $\lambda_N$ of the differential second-order
selfadjoint operator $A=-\nabla ^\mu\nabla _\mu+M^2+\xi R$. The eigenvalue
equation $A \Psi_N=\lambda_N \Psi_N$ can easily be solved for every value of 
$k$ in eq.~(\ref{gen-eucl-1}) by using the ansatz 
\begin{eqnarray}
\Psi_N=e^{-i\omega}\prod_{j=1}^{p}g_k(\rho_i)e^{il_j\varphi_j}.
\end{eqnarray}
In this relation, $l_i$ are integers and every function $g_j$ is a solution
of the equation 
\begin{eqnarray}  \label{gen-g}
\frac{1}{f_k(m\rho_j)}\frac{d}{d\rho_j}\big(f_k(m\rho_j)\frac{dg_j}{d\rho_j}%
\big) -\frac{m^2}{f_k^2(m\rho_j)} \left(l_j+\frac{4a\omega}{m^2}f_k^2(\frac{%
m\rho_j}{2})^2\right) g_j=\beta_jg_j,
\end{eqnarray}
where $\sum_j \beta_j=M^2+\xi R+\omega^2-\lambda_N$, the functions $g_j$
satisfying certain boundary conditions at the limits of the $\rho_j$
intervals. 

One consider the series with $s\in C$ (the prime on the sum means that any
possible null eigenvalues are omitted) 
\begin{eqnarray}  \label{zeta-def}
\zeta (s|A)={\sum_{N}}^{\prime }\lambda _{N}^{-s}.
\end{eqnarray}%
As is well-known, this series converges provided Re~$s>D/2$. It is possible
to continue the above sum into a meromorphic function of $s$ that is regular
at $s=0$ \cite{eli2}. In a path integral approach, the effective action for
a scalar field can be formally expressed as the functional determinant of
the operator $A$ as 
\begin{eqnarray}  \label{defIeff}
I_{eff}=-{\frac{1}{2}}\ln \det (A/\mu ^{2}),
\end{eqnarray}%
where $\mu $ is an arbitrary renormalization mass scale coming from the
path-integral measure. This determinant however is a formally divergent
quantity and needs to be regularized.

In a zeta function renormalization framework, the regularized determinant
reads 
\begin{eqnarray}
\ln \det (A/\mu ^{2})=-\zeta ^{\prime }(0|A)-\zeta (0|A)\ln \mu ^{2}.
\end{eqnarray}%
We note that since $\zeta (0|A)=0$ in odd dimensions (which is our case),
the dependence on the renormalization scale drops out. Relatively simple
expressions for the vacuum expectation values of the field fluctuations and
energy momentum tensor can be written by using the local zeta function
approach \cite{Moretti:1997qn}.

The series (\ref{zeta-def}) can be evaluated relatively easy in the zero
temperature limit of $k=0$ solutions. For $k=0$ metrics, we can easily solve
the equation (\ref{gen-g}) imposing on $g_{j}(\rho _{j})$ the boundary
conditions of finiteness at the origin and at infinity, which implies the
existence of $p$ more quantum numbers $n_{i}$ (see, also, 
refs.~\cite{Figueiredo:ze, probes} for a study of the Klein-Gordon equation in a 
$k=0$ background with $p=1,2$). This leads to an eigenvalue expression 
\begin{eqnarray}
\lambda _{N}=\omega ^{2}+2a\Big(\omega m_{i}+|\omega m_{i}|+|\omega
|(2n_{i}+p)\Big)+c,
\end{eqnarray}%
where $(N=m_{i},n_{i},\omega )$, the index $i$ runs from $1$ to $p$, and we
set $c=M^{2}+\xi R$, with the Ricci scalar $R=-2pa^{2}$. The massless
conformally coupled case corresponds to $c=-a^{2}(2p-1)/4$. We note the
existence of negative eigenvalues over some range of $\left( \xi ,M\right) $%
, which indicates that the ground state may be unstable. The functions $%
g_{0}(\rho _{i})$ in the general eigenvalue expression (\ref{gen-g}) are
given by 
\begin{eqnarray}
g_{0}(\rho _{i})=\rho _{i}^{|m_{i}|}e^{-\frac{|\omega |a\rho _{i}^{2}}{2}%
}L_{|m_{i}|}^{n_{i}}(|\omega |a\rho _{i}^{2}),
\end{eqnarray}%
with $L_{|m|}^{n}$ the Laguerre polynomial. Thus the zero-temperature zeta
function expression is \footnote{%
Since the volume of the spacetime diverges, we work here with scalar zeta
functions per unit volume.} 
\begin{eqnarray}
\zeta (s)=2\sum_{i=1}^{p}\int_{0}^{\infty }d\omega \sum_{m_{i}=-\infty
}^{\infty }\sum_{n=0}^{\infty }\Big(\omega ^{2}+2a\omega \big(%
m_{i}+|m_{i}|+2n_{i}+p\big)+c\Big)^{-s},
\end{eqnarray}%
the finite temperature counterpart corresponding to a multidimensional
inhomogeneous Epstein-type zeta function with a truncated range \cite%
{Elizalde:1997jv}.

We start by discussing the minimally coupled, massless case $(c=0)$. Here,
when evaluating the integral over $\omega $, we consider a cutoff $\epsilon $
at low $\omega $, eliminating the contribution of the $\omega =0$ null
eigenvalue. The apparent divergent integral as $\epsilon \rightarrow 0^{+}$
can be made harmless as in refs.~\cite{Hawking:1976ja,Moretti:1997qn}, \emph{%
after} one has fixed $\mathrm{{Re}~s}$ large finite, executed the
integration and performed the analytic continuation of this result at $s=0$.
The integral over $\omega $ is evaluated using the relation \cite{grad} 
\begin{eqnarray}
\int_{u}^{\infty }\frac{x^{\mu -1}dx}{(1+\beta x)^{\nu }}=\frac{u^{\mu -\nu }%
}{\beta ^{\nu }(\nu -\mu )}{}_{2}F_{1}(\nu ,\nu -\mu ;\nu -\mu +1;-\frac{1}{%
\beta u}),~~\mathrm{{Re~}\mu <{Re~}\nu ,}
\end{eqnarray}%
where ${}_{2}F_{1}(a,b;c;z)$ is the hypergeometric series. Applying the
quadratic transformation formula \cite{grad} 
\begin{eqnarray}
{}_{2}F_{1}(\alpha ,\beta ;2\beta ;z)=(1-\frac{z}{2})^{-\alpha }{}_{2}F_{1}(%
\frac{\alpha }{2},\frac{\alpha +1}{2};\frac{\beta }{2};(\frac{z}{2-z})^{2}),
\end{eqnarray}%
we find 
\begin{eqnarray}
\zeta (s)=\frac{(2a)^{1-2s}}{s-\frac{1}{2}}I(s)_{{}}\Xi ,
\end{eqnarray}%
where 
\begin{eqnarray}
I(s)=(1+2\epsilon )^{1-2s}{}_{2}F_{1}(s-\frac{1}{2},s;s+\frac{1}{2};\frac{1}{%
(1+2\epsilon )^{2}})
\end{eqnarray}%
is a regular function of $s$, which converges for $\epsilon >0$, and
\begin{eqnarray}  \label{xi}
\Xi =\sum_{i=1}^{p}\sum_{m_{i}=-\infty }^{\infty }\sum_{n=0}^{\infty }\Big(%
\frac{m_{i}+|m_{i}|}{2}+n_{i}+\frac{p}{2}\Big)^{1-2s}.
\end{eqnarray}
A regularized expression for (\ref{xi}) is obtained in several steps.
This
sum can be rewritten as
\begin{eqnarray}
\label{sd1}
\Xi =
\sum_{j=0}^{p}
(-\frac{1}{2})^{p-j}{p \choose j}
\sum_{m_i=0}^{\infty}
\sum_{n_i=0}^{\infty}
\Big(  m_1 +\dots + m_j  +
  n_1+\dots +n_p +\frac{p}{2}  \Big)^{1-2s},
\end{eqnarray}
(we replaced here $\sum_{m_{i}=-\infty }^{-1}(m_{i}+|m_{i}|+\alpha )=\zeta
_{R}(0)\alpha $).
The sum over $m_{i},n_{i}$ entering (\ref{sd1}) can be simplified by using
the relation \cite{Elizalde:fg} 
\begin{eqnarray}  \label{elizalde}
\sum_{n_1,\dots n_N=0}^{\infty}\left[c_1(n_1+\dots+n_N)+c\right]^{-s} =
c_1^{-s}\sum_{l=0}^{\infty} \left( {{N+l-2 }{l }} \right) \zeta_H(s,c/c_1+p),
\end{eqnarray}
where $\zeta _{H}(s,a)$ are the Hurwitz zeta functions, which are
meromorphic functions with a unique simple pole at $s=1$. However, an
evaluation of $\Xi $ may be obtained most easily using its representation in
terms of the Barnes zeta function \cite{barnes}
\begin{eqnarray}
\zeta _{\mathcal{B}}(s,a|\vec{r})=\sum_{\vec{m}=0}^{\infty }\frac{1}{(a+\vec{%
m}.\vec{r})^{s}},
\end{eqnarray}%
valid for $\mathrm{{Re~}s>d}$, with the $d$-vectors $\vec{m}$ and $\vec{r}$.
It is possible to continue the above sum into a meromorphic function of $s$,
the only poles of $\zeta _{\mathcal{B}}(s,a|\vec{r})$ being for $s=1,\dots d$%
, with
\begin{eqnarray}
\mathrm{Res\,\,}\zeta _{\mathcal{B}}(n,a|\vec{r})=\frac{%
(-1)^{d+n}B_{d-n}^{(d)}(a|\vec{r})}{(n-1)!(d-n)!\prod_{j=1}^{d}r_{i}},
\end{eqnarray}%
while
\begin{eqnarray}
\zeta _{\mathcal{B}}(-k,a|\vec{r})=\frac{(-1)^{d}k!B_{d+k}^{(d)}(a|\vec{r})}{%
(d+k)!\prod_{j=1}^{d}r_{i}},
\end{eqnarray}%
with $p=0,-1,\dots ,$ and $B_{a}^{(b)}$ the generalized Bernoulli
polynomials.

We find in this way the simple general expression
\begin{eqnarray}
\label{zetagen}
\zeta^{D=2p+1}(s)&=& \frac{(2a)^{1-2s}}{ s-\frac{1}{2}}I(s)_{}
\sum_{j=0}^{p}
(-\frac{1}{2})^{p-j}{p \choose j}
\zeb (2s-1,\frac{p}{2}|\vec 1_{p+j} )
\\
\nonumber
&=&
\frac{(2a)^{1-2s}}{ s-\frac{1}{2}}I(s)_{}
\sum_{j=0}^{p}
(-\frac{1}{2})^{p-j}{p \choose j}
\sum _{l=0}^{\infty}{p+j+l-2 \choose l} \zeta_H(2s-1,l+\frac{p}{2}).
\end{eqnarray}
The above expression can be simplified for particular values of $p$. A
useful relation in this context is \cite{Elizalde:fg}
\begin{eqnarray}
\sum_{m,n=0}^{\infty }[m+n+a]^{-s}=\zeta _{H}(s-1,a)-(a-1)\zeta _{H}(s,a).
\end{eqnarray}%
The results we find for $p=1,2$ are
\begin{eqnarray}
\zeta ^{D=3}(s) &=&\frac{(2a)^{1-2s}}{s-\frac{1}{2}}I(s)_{{}}\zeta _{H}(2s-2,%
\frac{1}{2})=\frac{(2a)^{1-2s}}{s-\frac{1}{2}}I(s)_{{}}(2^{2s-2}-1)\zeta
_{R}(2s-2),  \label{zeta4} \\
\zeta ^{D=5}(s) &=&\frac{(2a)^{1-2s}}{s-\frac{1}{2}}\frac{I(s)}{12}(\zeta
_{H}(2s-2,1)+2\zeta _{H}(2s-4,1))=\frac{(2a)^{1-2s}}{s-\frac{1}{2}}\frac{I(s)%
}{12}(\zeta _{R}(2s-2)+2\zeta _{R}(2s-4)).  \nonumber
\end{eqnarray}%
We observe that these series satisfy the standard relation $\zeta (0)=0$ and
present simple poles at $s=3/2,1/2~(p=1)$ and $s=5/2,3/2,1/2~(p=2)$, in
agreement with the general theory. \newline
The derivatives of these functions evaluated at $s=0$ are
\begin{eqnarray}
\zeta ^{\prime }{}^{D=3}(0)=6a\zeta _{R}^{\prime }(-2),~~\zeta ^{\prime
}{}^{D=5}(0)=-\frac{2a}{3}(\zeta _{R}^{\prime }(-2)+2\zeta _{R}^{\prime
}(-4)).
\end{eqnarray}%
%
%
%
The zeta function for the massive nonminimally coupled case can be
constructed following the result (\ref{zetagen}). By using an expansion in
terms of the parameter $c$, we find
\begin{eqnarray}
\label{zetagen-c}
\zeta^{D=2p+1}_c(s)&=& \frac{2(2a)^{1-2s}}{\Gamma(s)}
\sum_{k=0}^{\infty}(-1)^k(\frac{c}{a^2})^k\frac{1}{k!}
\frac{\Gamma(s+k)}{2s+2k-1}I(s+k)
\\
\nonumber
&_{}&\times
\sum_{j=0}^{p}
(-\frac{1}{2})^{p-j}{p \choose j}
\zeb (2s+2k-1,\frac{p}{2}|\vec 1_{p+j} ).
\end{eqnarray}
These zeta functions are analytic throughout the $s-$complex plane except
for $s=D/2-n$ (with $n=0,1,\dots )$ where simple poles appear.

Based on the above results, we can easily compute the effective action for a
scalar field with $M^2+\xi R\geq 0$. Unfortunately, we cannot use eq.~(\ref%
{zetagen-c}) in the physically more interesting massless conformally coupled
case, since $c<0$ in this limit and both null and negative eigenvalues $%
\lambda_N<0$ are present.
However, the zeta function approach can be
generalized to the case of negative eigenvalues
(see, $e.g.$, refs.~\cite{Vassilevich:2003xt, Adams:1995ej}).
In this case, the sum (\ref{zeta-def})
is taken over $|\lambda_{N}|^{-s}$,
a nontrivial phase factor
also appearing \cite{Adams:1995ej}.

It can easily be seen that, for the massless conformally coupled case, the $%
a $ dependence of the zeta function factorizes and it is similar to the
massless case, $\zeta (s)\sim a^{1-2s}$, the sum over $N=(n_{i},m_{i},\omega
)$ giving a function of $s$ only, whose evaluation requires a separate
discussion. Therefore, we expect the derivative of this function evaluated
at $s=0$ (thus the effective action of a massless conformally coupled scalar
field) to be proportional to $a$.


\end{document}